\documentclass[11pt, onecolumn]{IEEEtran}

\usepackage{graphicx} 
\usepackage{amsmath} 
\usepackage{amssymb}  
\usepackage{algorithm}
\usepackage{algorithmic}
\usepackage{cite}

\newtheorem{theorem}{Theorem}
\newtheorem{lemma}{Lemma}
\newtheorem{defn}{Definition}
\newtheorem{corollary}{Corollary}

\newcommand{\sth}[1]{\left\{ #1 \right\}}
\newcommand{\pth}[1]{\left( #1 \right)}
\newcommand{\qth}[1]{\left[ #1 \right]}

\newcommand{\prob}[1]{\Pr\left[#1\right]}
\newcommand{\Expect}{\mathbb{E}}
\newcommand{\expect}[1]{\mathbb{E}\left[#1\right]}
\newcommand{\indc}[1]{{\mathfrak{1}\left\{{#1}\right\}}}

\newcommand{\reals}{\mathbb{R}}

\newcommand{\bfY}{{\mathbf{Y}}}
\newcommand{\bfZ}{{\mathbf{Z}}}
\newcommand{\calC}{{\mathcal{C}}}
\newcommand{\calE}{{\mathcal{E}}}
\newcommand{\calM}{{\mathcal{M}}}
\newcommand{\calN}{{\mathcal{N}}}
\newcommand{\calP}{{\mathcal{P}}}
\newcommand{\calQ}{{\mathcal{Q}}}
\newcommand{\calT}{{\mathcal{T}}}
\newcommand{\calW}{{\mathcal{W}}}
\newcommand{\calX}{{\mathcal{X}}}
\newcommand{\calY}{{\mathcal{Y}}}
\newcommand{\calZ}{{\mathcal{Z}}}

\IEEEoverridecommandlockouts

\begin{document}

\title{Universal Clustering via Crowdsourcing}

\author{
Ravi Kiran Raman and Lav R.\ Varshney%
\thanks{This work was supported in part by NSF Grants IIS-1550145 and CCF-1623821,
and by Air Force STTR Contract FA8650-16-M-1819.}
\thanks{The authors are with the University of Illinois at Urbana-Champaign.}
}

\maketitle

\begin{abstract}
Consider unsupervised clustering of objects drawn from a discrete set, through the use of human intelligence available in crowdsourcing platforms. This paper defines and studies the problem of \emph{universal clustering} using responses of crowd workers, without knowledge of worker reliability or task difficulty. We model stochastic worker response distributions by incorporating traits of memory for similar objects and traits of distance among differing objects. We are particularly interested in two limiting worker types---temporary workers who retain no memory of responses and long-term workers with memory. We first define clustering algorithms for these limiting cases and then integrate them into an algorithm for the unified worker model. We prove asymptotic consistency of the algorithms and establish sufficient conditions on the sample complexity of the algorithm. Converse arguments establish necessary conditions on sample complexity, proving that the defined algorithms are asymptotically order-optimal in cost.
\end{abstract}


\section{Introduction}
\label{sec:intro}
Crowdsourcing has grown in recent times as a potent tool for performing complex tasks using human skill and knowledge. It is increasingly being used to collect training data for novel machine learning problems. Almost \emph{a fortiori}, there is no prior knowledge on the nature of the task and so the use of general human intelligence has been needed \cite{KitturCS2008}. As such, this setting requires processing human-generated signals in the absence of prior knowledge about their properties; this setting requires universality.

Because crowdsourcing often employs unreliable workers, the signals they generate are inherently noisy \cite{IpeirotisPW2010}. Hence, responses of crowd workers are modeled as outputs of a noisy channel. Although these channels are unknown to the employer, crowdsourcing techniques have thus far made assumptions on either the channel distribution or structure to design appropriate decoders. We define an alternative approach---\emph{universal crowdsourcing}---that designs decoders without channel knowledge, and develop achievability and converse arguments that demonstrate order-optimality.

The emergence of diverse online crowdsourcing platforms such as Amazon Mechanical Turk and Upwork (formerly oDesk) has created the option of choosing between temporary workers and long-term workers. That is, tasks can be completed by either soliciting responses from a large number of workers performing small parts of a large task, or a specialized group of employees who work long-term on the task at hand. 

The trade-off between reliability and cost of each type of worker pool warrants systematic study. Whereas temporary workers are inexpensive and easily available, some labor economists argue the excess cost of long-term employment is worthwhile due to the reliability and quality of work it ensures \cite{Scheiber2015}. However, no quantitative characterization for this conjecture exists.  As we will see, the results of this paper allow such comparisons.

Since workers are human, they are subject to several factors that affect human decision making, as identified in behavioral economics \cite{GinoP2008}.

A standard assumption in crowdsourcing and in universal clustering \cite{KargerOS2014,MisraW2013} has been independent and identically distributed (i.i.d.) worker responses across time/tasks. However, empirical evidence argues against temporal independence for individual worker responses \cite{JungPL2014}.
Due to the \emph{availability heuristic}, worker responses may rely on the immediate examples that come to a person's mind, indicating memory in responses across tasks/time. 
Further, this influence may be more due to salient (vivid) information rather than full statistical history. 

Due to the \emph{anchoring and adjustment heuristic}, people tend to excessively rely on a specific trait of an object in decision making and further due to the \emph{representativeness heuristic}, people tend to assume commonality among objects. These traits indicate there is a notion of distance among the response distributions corresponding to different objects.

To capture memory and distance, we define a unified model of worker responses, and then study two limiting cases. First we consider responses of temporary workers who respond independently across tasks and across workers. We then consider responses of long-term workers with object-specific memory. Specifically, we consider a Markov memory model wherein the response to an object is dependent on the most recent response and the response to the most recent occurrence of an object of the same class; generalizations to other Markov models follows readily. In both cases, we address questions of universality and sample complexity for reliable clustering, providing benchmarks for worst-case performance.

\subsection{Prior Work}

There is a vast and rich literature on crowdsourcing and clustering; we describe a non-exhaustive listing of particularly relevant prior work.

Algorithm design for crowdsourcing typically focuses on minimizing the cost of reliability. In particular, algorithms with order-optimal budget-reliability trade-offs have been designed for binary classification with unknown (but i.i.d.) crowd reliabilities \cite{KargerOS2014}. Efficient algorithms for multi-class labeling have been proposed in \cite{KargerOS2013} albeit without cost optimality guarantees. More recently, non-parametric permutation models of crowd workers were considered for binary classification \cite{ShahBW2016_arxiv}.

An alternate strategy for multi-class labeling with workers lacking sufficient domain expertise is to decompose the overall task into simpler subtasks, and introduce redundancy through an error control code \cite{VempatyVV2014}. This approach is implicitly effective for mismatched crowdsourcing for speech transcription \cite{VarshneyJH2016}.

Separate from crowdsourcing, the problem of clustering has been widely studied. Algorithms such as $k$-means clustering and its generalization to other Bregman divergence similarity measures \cite{BanerjeeMDG2005} are popular methods that incorporate distance-based clustering.

The problem of universal clustering was considered in a communication setting \cite{MisraW2013,Misra2015}, such that messages communicated across an unknown channel, after encoding using a random codebook, are clustered by exploiting dependency among outputs of similar messages. Particularly the decoder uses the minimum partition information functional \cite{ChanAEKL2015} to perform optimal clustering. Similar information-based agglomerative clustering schemes have also been explored \cite{SlonimFT2002}.

Classification using crowdsourced responses in a clustering framework, followed by a labeling phase performed by a domain expert has been studied experimentally \cite{ZhangSWW2016}. 

\subsection{Main Contributions}

In this work, we provide a theoretical study of universal crowdsourcing for clustering. In particular, we focus on the design of universal clustering algorithms with provable asymptotic consistency and order optimality in sample complexity. 

The presence of memory in worker response demands an approach that differs from past crowd algorithms defined for i.i.d.\ models \cite{KargerOS2014}. Notwithstanding \cite{VempatyVV2014}, in the crowdsourcing framework herein, we do not have the opportunity to encode messages. This calls for a treatment different from past work in universal clustering \cite{MisraW2013}.

We first consider the case of temporary workers without memory, wherein, we design a distance-based universal clustering decoder that uses distributional identicality among objects of the same class. That is, two objects are clustered together if the $f$-divergence between the conditional distributions of the responses is small. We prove asymptotic consistency of the algorithm and prove order optimality in sample complexity. The algorithm applies directly to a large class of similarity measures.

We then consider long-term employees with object-specific memory. Specifically we consider a Markov memory model wherein the response to an object is dependent on the most recent response and the response to the most recent occurrence of an object of the same class. For this model, we show the existence of information-based universal clustering strategies that perform asymptotically reliable clustering. Further, we study the sample complexity of the decoder and show order optimality by comparing with the necessary cost. We also highlight the extension of the algorithm to higher-order Markov memory models. We also observe that the universal clustering algorithm is structurally similar to traditional clustering algorithms such as the MIRN (mutual information relevance networks) clustering \cite{NaganoKI2010}, and the minimum partition information clustering \cite{ChanAEKL2015} algorithms under added constraints on the channel model.

Finally, we use results obtained for these two limiting cases to construct a clustering algorithm for a unified worker model. We prove asymptotic consistency. Further, we show order optimality of the sample complexity as a function of the number of objects to be clustered.

\section{Model} \label{sec:model}
\subsection{System}
We formalize the model of the crowdsourcing system by first describing a unified worker model and then specializing to the cases of workers with and without memory. We aim to design universal decoders that cluster a given set of objects using the crowd responses. For any index vector $S$, let $Z_s$ be the set $\{Z_i:i\in S\}$, where $Z$ is any vector. Similar notation shall be used for a matrix of indices $S$ and matrix $\bfZ$. 

\begin{figure}[t]
	\centering
	\includegraphics[scale=0.5]{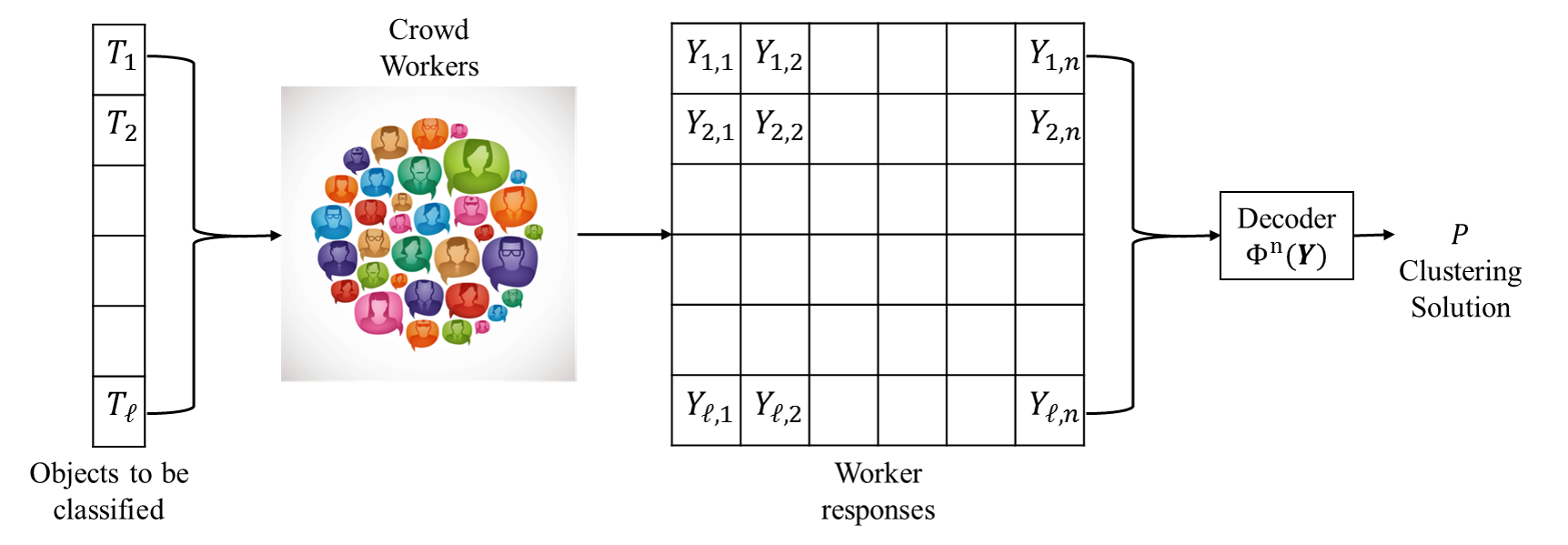}
 	\caption{Model of the crowdsourcing system}
  	\label{fig:crowd_model}
\end{figure}

We consider the problem of crowd workers employed to perform classification of objects. For instance, consider the task of classifying images of dogs according to their breeds. The workers observe images and respond with the breed of the dog in the image. Since worker responses are noisy, in the absence of knowledge of worker channels it is not feasible to identify the labels (breeds of dogs) accurately. Thus, we aim to cluster the dogs according to their breeds and determine the labels of each cluster by using a domain expert. The crowdsourcing system model is depicted in Fig.~\ref{fig:crowd_model}.

Let $\calT$ be a finite alphabet of object clusters. Without loss of generality, let us assume that $\calT = [\tau] = \{1,\dots,\tau\}$, where $\tau < \infty$ is a constant. Each object is viewed by crowd workers as $X \in \calX$, which has some relation to its type $T \in \calT$. 

Let the set of objects to be clustered be $\sth{X_1,\dots,X_{\ell}}$. Let the label of object $X_i$ be $T_i$, for all $i \in [\ell]$. That is, the objects to be clustered are treated as manifestations of the various object labels. Thus clustering the set of objects is the same as clustering according to their object labels. Let us assume that the objects are drawn according to an unknown prior $P_T(\cdot)$ on the set of classes.

For each object $X_i, i \in [\ell]$, the crowdsourcing system solicits $n$ responses, $Y_i^n$, from workers employed to classify the objects according to their labels. The collection of responses is given by the matrix $\bfY = \pth{ Y^n_1,\dots,Y^n_\ell }' = \qth{Y_{i,j}}_{i \in [\ell], j \in [n]}$. Let $W \in \calW$ be the index of a worker and let $S_W \subset [\ell]\times[n]$ be the index set corresponding to the responses offered by worker $W$ in $\bfY$. We assume that $S_w \cap S_{\hat{w}} = \emptyset$ for any two workers $w \neq \hat{w}$ and $\cup_{w\in\calW} S_w = [\ell]\times[n]$.

Let $\calQ$ be the set of all conditional probability mass functions (pmfs) that characterize worker responses. Then, $Y_{S_W} \sim Q^{(W)}(Y_{S_W} \vert T_{\hat{S}_W})$ where $Q^{(w)}(\cdot) \in \calQ$ is the distribution characterizing the response of worker $w \in \calW$ and $\hat{S}_W$ is the corresponding set of indices of objects. Models of these distributions are detailed later.

With regard to the example of classifying dog images, the object label, $T$, is the breed of the dog in an image; the object, $X$, is the image of a particular dog; and the responses of the crowd workers $Y^n$ are the breeds they categorize the image to. 

Since clustering is performed solely based on the response of the workers, it is fair to assume that the number of clusters that can be formed is directly dependent on $|\calY|$. However, it is not essential for every worker to answer every question in a practical crowdsourcing platform. Thus we assume that the workers either respond with an answer in $\calT$ or offer a `null' response $\xi$ to every task \cite{LiVVV_arxiv}. Thus, without loss of generality, we assume that $Y_{ij} \in \calY = \calT \cup \{\xi\}$, for all $i \in [\ell], j \in [n]$.

\subsection{Universal Clustering Performance}

\begin{defn}[Correct Clustering] \label{defn:opt_clust}
A \emph{clustering} of a set of objects $X_1,\dots,X_{\ell}$ is a partition $P$ of $[\ell]$. The sets of a partition are referred to as \emph{clusters}. The clustering is said to be \emph{correct} if
\[
T_i,T_j \in C \iff T_i = T_j,
\]
for all $i,j \in [\ell]$, $C \in P$. For a given set of object labels, $T^\ell$, let $P^{*}\pth{T^\ell}$ be the correct clustering. 
\end{defn}
Let $\calP$ be the set of all partitions of $[\ell]$.

\begin{defn}[Partition Ordering]
A partition $P$ is \emph{finer} than $P'$, if the following ordering holds
\begin{equation*}
P \preceq P' \iff \text{ for all } C\in P, \text{ there exists } C'\in P' : C \subseteq C'.
\end{equation*}
Similarly, a partition $P$ is said to be \emph{denser} than $P'$ if $P \succeq P' \iff P' \preceq P$.
\end{defn}

\begin{defn}[Universal Clustering Decoder]
A \emph{universal clustering decoder} is a sequence of functions $\Phi^{(n)}: \bfY \rightarrow \calP$ that are designed in the absence of knowledge of $\calQ$ and $P_T$. Here the index $n$ corresponds to the number of crowd responses collected per object.	
\end{defn}

We now define how to characterize decoder performance.
\begin{defn}[Error Probability]
Let $\Phi^{(n)}(\cdot)$ be a universal decoder. Then, the \emph{error probability} is given by
\begin{align}
P_e(\Phi^{(n)}) &= \prob{\Phi^{(n)}(\bfY) \neq P^*\pth{T^{\ell}}} \nonumber \\
&= \Expect_{P_T^{\otimes \ell}} \qth{\expect{ \indc{\Phi^{(n)}(\bfY) \neq T^\ell} \vert T^{\ell}}}, \label{eqn:clustering_error}
\end{align} 
where $P_T^{\otimes \ell} (t^\ell) = \prod_{i=1}^{\ell} P_T(t_i)$ and $\indc{\cdot}$ is the indicator function.
\end{defn}

\begin{defn}[Asymptotic Consistency]
A sequence of decoders $\Phi^{(n)}$ is said to be \emph{universally asymptotically consistent} if
\begin{equation} \notag
\lim_{n \rightarrow \infty} P_e(\Phi^{(n)}) = 0, \mbox{ for all } P_T \in \calM(\calT),
\end{equation}
where $\calM(\cdot)$ is the space of all prior distributions on the set of objects, $\calT$. 
\end{defn}

\begin{defn}[Sample Complexity]
Let $\epsilon > 0$ be the permissible error margin. Then the \emph{sample complexity} of the universal clustering problem is
\[
N^*(\epsilon) = \min \sth{n \in \mathbb{N} : \max_{P_T \in \calM(\calT)} P_e(\Phi^{(n)}) < \epsilon },
\]
where the minimum is taken over the set of all sequences of universal decoders $\Phi^{(n)} $.
\end{defn}

For simplicity, we will use $\Phi$ to denote $\Phi^{(n)}$ when it is clear from context. 

\subsection{Workers}
We now define a model to characterize crowd worker responses. Let us assume the crowdsourcing system employs $n$ crowd workers chosen at random from $\calW$. Without loss of generality, let us assume that the set of workers chosen is $[n]$. We assume that each worker responds to every object $X_i, i \in [\ell]$ and the set of responses of worker $j$ is $\{Y_{1,j},\dots,Y_{\ell,j}\}$. We assume that the responses of each worker $w \in \calW$ are drawn according to the conditional distributions $Q^{w}$.

We assume that the responses of each worker depend on prior responses in a Markov sense. In particular, define the set of neighbors as $\calN_i = \sth{\{i-1\} \cup \{\max \{k \in [i-1] : T_k = T_i\}\}}$ for any $i \in [\ell]$ i.e., the most recent object and the most recent occurrence of an object of the same class. The response of any worker $j \in [n]$, for any $k \leq \ell$ satisfies
\begin{align}
\label{eqn:memory_kernel}
&\prob{(Y_{1,j},\dots,Y_{k,j}) = (y_1,\dots,y_k) \vert T^{k} = t^{k}} = \prod_{i=1}^k Q^{(j)}(y_i \vert y_{\calN_i},t_i). 
\end{align}

Additionally we assume that for any worker $j \in [n]$ and $i \leq \ell$, for every $t \in \calT$,
\begin{equation}
\label{eqn:unified_kernel}
\prob{Y_i = y \vert T_i = t} = Q^{(j)}\pth{y \vert t}.
\end{equation}
That is, the worker responses are dependent on the prior responses such that the marginal conditional distribution of the response, given an object is identical across objects of the same class (that is, the marginals are invariant across permutations of the given set of objects).

We also assume that, given the object, the responses are independent across workers. That is,
\begin{align} 
\label{eqn:worker_indep_kernel}
&\prob{ (Y_{i,1},\dots,Y_{i,k})=(y_1,\dots,y_k) \vert T_i=t } = \prod_{j=1}^k \prob{ Y_{i,j} = y_j \vert T_i= t}.  
\end{align} 
Thus, the unified worker model of crowd responses is characterized by \eqref{eqn:memory_kernel}, \eqref{eqn:unified_kernel}, and \eqref{eqn:worker_indep_kernel}.

As mentioned in Section~\ref{sec:intro}, in addition to the unified worker model, we focus on two special classes of workers---temporary workers without memory and long-term workers with memory.

In particular, when temporary workers are employed we assume they do not retain memory of their prior responses and so are independent across time. In order to solicit such responses, we may assume that the crowdsourcing platform delegates each task to a sequence of $n$ workers selected uniformly at random from a sufficiently large crowd to ensure independence of responses across objects. That is, if the index of responses of worker $w$ is $S_w \subset [\ell] \times [n]$, then
\begin{align}
\prob{\bfY_{S_w} = y_{S_w} \vert T^{\ell} = t^{\ell}} = \prod_{(i,j) \in S_w} Q^{(w)}(y_{i,j} \vert t_i). \label{eqn:temp_kernel}
\end{align}
Further, the responses also satisfy \eqref{eqn:worker_indep_kernel}.

The second special class of workers we consider is long-term workers with Markov memory. In particular, responses of this class of workers are characterized by \eqref{eqn:memory_kernel} and \eqref{eqn:worker_indep_kernel}.

\section{Temporary Workers}

Consider the scenario of clustering using responses of temporary workers (without memory). As mentioned in the system model, we assume worker responses are independent across objects and workers. 

We first introduce $f$-divergences and their properties, so as to characterize the quality of crowd responses. We then define the universal clustering algorithm and prove asymptotic consistency and order optimality in sample complexity.

\subsection{$f$-Divergence}

To measure the separation among the conditional distributions of crowd responses to different object classes, we use the  Csisz{\'a}r $f$-divergence \cite{AliS1966,Csizar1967}.
\begin{defn}[$f$-divergence]
Let $p,q$ be discrete probability distributions defined on a space of $m$ alphabets. Given a convex function $f:[0,\infty) \rightarrow \reals$, the \emph{$f$-divergence} is defined as:
\begin{equation} \label{eqn:f-div_defn}
D_f(p\|q) \triangleq \sum_{i=1}^m q_i f\pth{\frac{p_i}{q_i}}.
\end{equation}
The function $f$ is said to be \emph{normalized} if $f(1)=0$. 
\end{defn}

Some specific $f$-divergences are the KL divergence $D(p\|q)$ and the total variational distance $\delta(p,q)$. Specifically, the KL divergence and the total variational distance are the $f$-divergences corresponding to the functions $f(x) = x \log x$ and $f(x) = |x-1|$ respectively. We now state some bounds for $f$-divergences. 

\begin{theorem}[{\cite[Chapter II.1]{Dragomir2000}}] \label{thm:f-div_bound}
Let $p,q$ be discrete probability distributions on a space of $m$ alphabets such that there exist $r,R$ satisfying $0 \le r \le \frac{p_i}{q_i} \le R \le \infty$ for all $i \in [n]$. Let $f:[0,\infty) \rightarrow \reals$ be a convex and normalized function satisfying the following criteria:
\begin{enumerate}
\item $f$ is twice differentiable on $[r,R]$, and
\item there exist real constants $c,C<\infty$ such that
\begin{equation*}
c\le xf''(x)\le C, \text{ for all } x\in(r,R).
\end{equation*}
\end{enumerate}
Then, we have,
\begin{equation} \label{eqn:f-div_bound}
cD(p\|q)\le D_f(p\|q)\le CD(p\|q).
\end{equation} 
\end{theorem}

For ease, we refer to the constraints on $f$ in Theorem \ref{thm:f-div_bound} as \emph{smoothness} constraints. If $f$ is twice differentiable in $[r,R]$, then we know that there exists a constant $L$ such that $f$ is $L$-Lipschitz.
\begin{theorem}[{\cite[Chapter II.3]{Dragomir2000}}] \label{thm:f-div_upper}
Let $f:[0,\infty)\rightarrow\mathbb{R}$ be convex, normalized, and $L$-Lipschitz on $[r,R]$. Then,
\begin{equation}\label{eqn:f-div_upper}
0\le D_f(p\|q)\le L\delta(p,q).
\end{equation}
\end{theorem}
Further, Pinsker's inequality lower bounds the KL divergence $D(p\|q)$ with respect to the total variational distance as
\begin{equation} \label{eqn:Pinsker}
D(p\|q)\ge (2\log_2 e) \delta^2(p,q).
\end{equation}

\begin{corollary} \label{cor:f-div_tvd_bound}
For any convex and normalized function $f$ that satisfies the smoothness constraints and is $L$-Lipschitz,
\begin{equation}\label{eqn:f-div_tvd_bound}
\kappa\delta^2(p,q)\le D_f(p\|q)\le L\delta(p,q),
\end{equation}
where $\kappa=2c\log_2 e$.
\end{corollary}
\begin{IEEEproof}
The result follows from Theorems \ref{thm:f-div_bound} and \ref{thm:f-div_upper}, and \eqref{eqn:Pinsker}.
\end{IEEEproof}

These inequalities are used to prove consistency of the designed universal clustering decoder.

\subsection{Task Difficulty for Worker Pool}

Let $Q_1,\dots,Q_{\tau}$ be the conditional response distributions given the object class, defined as
\[
Q_i(y) \triangleq \prob{Y = y \vert T = i} = \expect{Q^{(W)} ( Y = y \vert T = i)},
\]
where the expectation is taken over the workers in the pool. Since the responses are obtained from temporary workers chosen at random, it suffices to consider these expected conditional response distributions.

\begin{defn}[Distance Quality] \label{defn:task_diff}
For a given pool of temporary workers, the \emph{difficulty of the tasks} is defined as
\begin{equation} \label{eqn:theta_dist_sep}
\theta_d \triangleq \min_{\{i,j \in \calT, i \neq j\}} D_f(Q_i\|Q_j),
\end{equation}
where $D_f$ corresponds to the $f$-divergence chosen as the notion of similarity for the problem at hand.
\end{defn}
The operational significance of this informational definition of distance (Definition \ref{defn:task_diff}) will emerge in coding theorems Lemma~\ref{lemma:temp_optimal_rest} and Theorem~\ref{thm:temp_consistency}.

Clustering is performed using the maximum likelihood estimates of the $f$-divergence between distributions corresponding to responses to objects. Convergence of the empirical estimates is asymptotically consistent and the rates of convergence are discussed in Appendix \ref{app:dist_conv}.

\subsection{Universal Clustering using Temporary Workers}

\begin{figure}[b]
	\centering
	\includegraphics[scale=0.75]{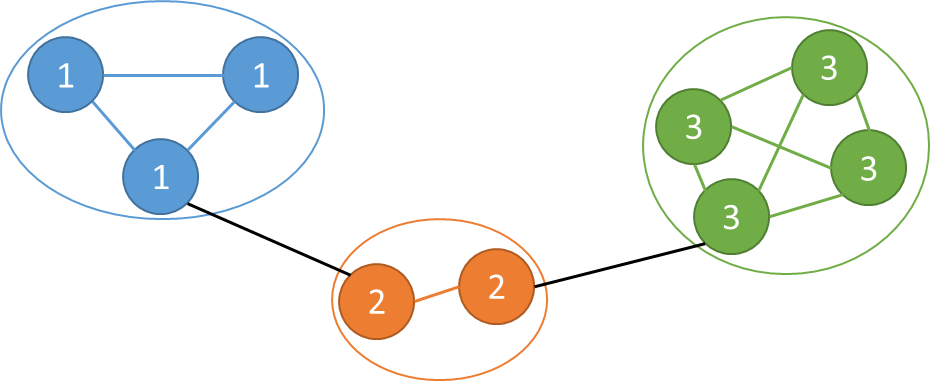}
 	\caption{Distance-based clustering of $9$ objects of $3$. The graph is obtained by thresholding the $f$-divergences of empirical distributions of responses to objects. The clustering is then done by identifying the maximal clusters in the thresholded graph.}
  	\label{fig:dist_cluster}
\end{figure}

Responses to objects of the same class are identical in distribution. Thus, we perform universal clustering, $\Phi_{\text{temp}}(\bfY)$, according to Algorithm \ref{algo:clustering}. That is, the algorithm identifies the cliques in the graph obtained by thresholding the $f$-divergence between the corresponding empirical distributions. The functioning of the algorithm is depicted in Fig. \ref{fig:dist_cluster}.
\begin{algorithm}[t]
  \caption{Clustering with temporary workers, $\Phi_{\text{temp}}(\bfY)$}
  \label{algo:clustering}
  \begin{algorithmic}[t]
   \IF {$f(x) = |x-1|$}
   	\STATE{$\gamma_n \leftarrow c_1 n^{-\alpha}$, where $c_1$ is a constant and $\alpha \in [0,1/2]$}
   \ELSE
   	\STATE{$\gamma_n \leftarrow c_1 n^{-\beta}$, where $c_1$ is a constant and $\beta \in [0,1]$}	
   \ENDIF
   \STATE {Determine empirical distributions $q_j$, $j \in [\ell]$}
   \STATE {Construct $G = ([\ell], E)$, s.t. $(i,j) \in E \iff D_f(q_i,q_j) \leq \gamma_n$} 
   \STATE {$\calC = \{C : C \text{ is a maximal clique in } G\}$}
   \STATE {Select minimal weight partition of $[\ell]$ from $\calC$}
  \end{algorithmic}
 \end{algorithm}

\begin{lemma} \label{lemma:temp_optimal_rest}
For $f(x) = |x-1|$ and $\gamma_n = c_1 n^{-\alpha}$, $\alpha \in [0,1/2]$ let $\gamma_n < \theta_d/2$. For any other convex function $f$ satisfying the smoothness constraints let
\[
\gamma_n = c_1 n^{-\beta} < \frac{\kappa}{2L} \theta_d^2 + 2\frac{\kappa}{L^2}\pth{1-\sqrt{1+\frac{L\theta_d^2}{2}}},
\]
where $\kappa = 2c\log_2 e$ and the function $f$ is $L$-Lipschitz.

Define the ball of radius $\rho$ centered at $p$ as
\[
B_f(p,\rho) = \sth{q: D_f(p\|q) \leq \rho}.
\]
If for all $i \in [\ell]$, the empirical distribution of responses $q_i \in B_f(Q_{t_i},\gamma_n/2)$ then, $\Phi_{\text{temp}} (\bfY) = P^*(t^\ell)$, the correct clustering of the set of objects.
\end{lemma}
\begin{IEEEproof}
Let us first consider $f(x) = |x-1|$. Since $q_i \in B_f(Q_{t_i},\gamma_n/2)$ for all $i \in [\ell]$, we have
\[
\max_{\{i,j\in [\tau]: T_i = T_j\}} \delta(q_i,q_j) \leq \gamma_n \leq \frac{\theta_d}{2} \leq \min_{\{i,j\in [\tau]: T_i \neq T_j\}} \delta(q_i,q_j).
\]

Let $C_i = \{ j \in [\ell]: T_j = i \}$, for any $i \in [\tau]$. Then, for $j,k \in C_i$, $D_f(q_j,q_k) < \gamma_n$ and so $(j,k) \in E$. Thus, $C_i$ is a clique of $G$. 

Further, this observation also implies that for any $i \in C_t , j \in C_{t'}, t \neq t'$, $D_f(q_i,q_j) > \gamma_n$ and so $(i,j) \notin E$. Thus, any set $C \subseteq [\ell]$ such that there exist $i,j \in C$, with $T_i \neq T_j$, is not a clique in $G$.

Thus $C_i$ is a maximal clique in $G$ for all $i \in [\tau]$. Thus $\Phi_{\text{temp}}(\bfY) = \{C_i: i \in [\tau]\} = P^*(t^{\ell})$, the correct partition.

For the second part of the lemma, from (\ref{eqn:f-div_tvd_bound}), we note that the condition on $\gamma_n$ guarantees
\[
\max_{\{i,j\in [\tau]: T_i = T_j\}} D_f(q_i\|q_j) \leq \gamma_n < \min_{\{i,j\in [\tau]: T_i \neq T_j\}} D_f(q_i\|q_j).
\]
Thus the result follows from a very similar argument.
\end{IEEEproof}

From Lemma \ref{lemma:temp_optimal_rest}, we observe that when the empirical distributions are sufficiently close to the corresponding true distributions, the algorithm outputs the correct clustering. Using this result, we prove consistency of the algorithm.

\begin{theorem} \label{thm:temp_consistency}
If $f(x) = |x-1|$, then for any $\alpha \in (0,1/2)$ and constant $c_1$, for
\begin{equation} \label{eqn:temp_TV_cost}
n \gtrsim \max\sth{\pth{\frac{2 c_1}{\theta_d}}^{1/\alpha} , \pth{\frac{4\log \ell}{c_0c_1^2}}^{1/(1-2\alpha)}},
\end{equation}
sufficiently large, $\Phi_{\text{temp}}(\cdot)$ achieves arbitrarily low clustering error probability. For fixed $\ell$ and $\theta_d$, it is universally asymptotically consistent.

For any other convex, normalized function $f$ satisfying the smoothness constraints, for any $\beta \in (0,1)$ and constant $c_1$, for
\begin{align} \label{eqn:temp_f_cost}
&n \gtrsim \max\sth{\pth{\frac{2c_1 L^2}{\kappa} \mu_{\theta_d}}^{1/\beta}, \pth{\frac{C \log l}{c_1}}^{1/(1-\beta)}},
\end{align}
where,
\[
\mu_{\theta_d} = \pth{L\theta_d^2 + 4\kappa\pth{1-\sqrt{1+\frac{L\theta_d^2}{2}}}^{-1}},
\]
with sufficiently large constant $c_1$, $\Phi_{\text{temp}}(\cdot)$ achieves arbitrarily low clustering error probability. For fixed $\ell$ and $\theta_d$, it is universally asymptotically consistent.
\end{theorem}

\begin{IEEEproof}
For $f(x) = |x-1|$ and $n \geq \pth{\frac{2 c_1}{\theta_d}}^{1/\alpha}$, $\gamma_n \leq \theta_d/4$. Thus, when $f(x) = |x-1|$, we can bound the error probability as follows
\begin{align}
P_e(\Phi_{\text{temp}}) &\leq \Expect_{P_T^{\otimes l}}\qth{\prob{\exists i \in [l]: q_i \notin B(Q_{t_i}, \gamma_n/2)} } \notag\\
&\leq \ell \qth{ (n+1)^{|\calY|} \exp \pth{-c_0 n\frac{\gamma_n^2}{4}} } \label{eqn:Sanov_Pinsker}\\
&= \exp \pth{\pth{ \log \ell + (\tau+1)\log (n+1) - \frac{c_0c_1^2}{4} n^{1-2\alpha}}} , \notag
\end{align}
where \eqref{eqn:Sanov_Pinsker} follows from the union bound and Lemma~\ref{lemma:emp_conc}. Thus, the cost conditions given in \eqref{eqn:temp_TV_cost} and asymptotic consistency follow.

Using a similar argument and Lemma \ref{lemma:temp_optimal_rest}, we obtain \eqref{eqn:temp_f_cost}.
\end{IEEEproof}

\begin{corollary} \label{cor:sample_comp_temp}
Given $\calT = [\tau]$ with $\tau < \infty$ a constant:
\begin{enumerate}
\item for a constant $\theta_d > 0$, $N_{\text{temp}}^*(\epsilon) = O\pth{(\log \ell)^{1/(1-\beta)}}$ and taking $\beta \rightarrow 0$, we observe that $N_{\text{temp}}^*(\epsilon) = O\pth{\log \ell}$ for any of the similarity metrics;
\item for a constant $\ell$, for $f(x) = |x-1|$, $N_{\text{temp}}^*(\epsilon) = O\pth{\theta_d^{-1/\alpha}}$ and taking $\alpha \rightarrow 1/2$, $N_{\text{temp}}^*(\epsilon) = O(\theta_d^{-2})$. On the other hand, for other convex functions $f$ satisfying the smoothness constraints, $N_{\text{temp}}^*(\epsilon) = O(\theta_d^{-1/\beta})$ and specifically, taking $\beta \rightarrow 1$, $N_{\text{temp}}^*(\epsilon) = O(\theta_d^{-1})$.
\end{enumerate}
\end{corollary}
\begin{IEEEproof}
The results follow directly from Theorem \ref{thm:temp_consistency}.
\end{IEEEproof}

To summarize, this subsection has defined a universal clustering algorithm that is asymptotically consistent and also described sufficient conditions on sample complexity of universal clustering.

\subsection{Lower Bound on Sample Complexity}

We now show matching lower bounds on the sample complexity for universal clustering.

\begin{theorem} \label{thm:temp_conv}
Let  $\min_{i,j \in [\tau]} \delta(Q_i,Q_j) = \theta_d$. Then the sample complexity of universal clustering satisfies
\[
N_{\text{temp}}^* = \Omega\pth{\frac{\log \ell - \log \epsilon}{\theta_d^{2}}}.
\]
\end{theorem}
\begin{IEEEproof}
Consider the prior distribution $P_T$ such that $P_T(1) = P_T(2) = 0.5$. Let $\psi_{ij}$ be the binary hypothesis testing problem given by:
\begin{equation} \label{eqn:binary_ij}
\psi_{ij}:\begin{cases}
H_0: T_i = T_j  \\
H_1: T_i \neq T_j
\end{cases}.
\end{equation}
There exists $\binom{\ell}{2}$ such binary hypothesis tests. Choose a set of tests $\tilde{\psi} \subset \{\psi_{ij}:i,j \in [\ell], i \neq j\}$ of cardinality $\ell/2$ such that no two tests in the set share a common object. This indicates that the binary hypothesis tests are independent of each other owing to the independence across objects.

Let $\Phi$ be a decoder for the clustering problem. Then a correct solution to the clustering problem implies a correct solution to $\psi_{ij}$, for all $i,j \in [\ell], i \neq j$. This implies that an instance of correct clustering translates to correct decisions in all tests in $\tilde{\psi}$. Thus,
\begin{align}
P_e(\Phi) &\geq 1 - \prod_{\{i,j \in [\ell]: \psi_{ij}\in \tilde{\psi}\}} \pth{1 - \prob{\text{error in }\psi_{ij}}} \label{eqn:binary_test_indep} \\
&\geq 1 - \prod_{\{i,j \in [\ell]: \psi_{ij}\in \tilde{\psi}\}} \pth{1 - \frac{1}{4}\exp\pth{-2nB_{ij}}} \label{eqn:Kailath_LB} \\
&\geq 1 - \pth{1 - \frac{1}{4}\exp\pth{-2nB_{\text{max}}}}^{\lfloor \ell/2 \rfloor} \label{eqn:min_B_ij} \\
&= \frac{\lfloor \ell/2 \rfloor}{4}\exp\pth{-2nB_{\text{max}}} + o\pth{\exp\pth{-4nB_{\text{max}}}}, \label{eqn:sample_LB}
\end{align}
where (\ref{eqn:binary_test_indep}) follows from the independence of the binary tests and (\ref{eqn:Kailath_LB}) follows from the Kailath lower bound \cite{Kailath1967}. Here $B_{ij}$ is the Bhattacharyya distance corresponding to the hypotheses of the test $\psi_{ij}$ and considering non-triviality, there exists a test such that $B_{ij}>0$. Thus bounding from below by the test with maximum distance (\ref{eqn:min_B_ij}), $B_{\text{max}} = \max_{\{i,j \in [\ell]: \psi_{ij}\in \tilde{\psi}\}} B_{ij} > 0$ and using the binomial expansion, we obtain (\ref{eqn:sample_LB}). 

Now, using Jensen's inequality, we have $B_{ij} \leq \frac{1}{2} \pth{D(Q_1 \| Q_2) + D(Q_2 \| Q_1)}$. This follows from the definition of the binary hypotheses tests and the independence of samples.

From Pinsker's inequality and reverse Pinsker's inequality \cite{CsiszarZ2006}, we have
\[
(2 \log_2 e) \delta^2(P,Q) \leq D(P\|Q) \leq \pth{\frac{4\log_2 e}{Q_{\text{min}}}} \delta^2 (P,Q),
\]
where $Q_{\text{min}} = \min_{x \in \text{supp} (P)} Q(x)$ and $\text{supp}(\cdot)$ is the support of the distribution. Since we are concerned with the sample complexity in the worst case when $\delta(P,Q) \rightarrow 0$, it suffices to consider $Q_{\text{min}} > 0$. Thus, the above bounds indicate that
\[
D(Q_1 \| Q_2) \asymp D(Q_2 \| Q_1) \asymp \theta_d^2,
\]
where it is said $g \asymp h$, if there exists constants $a,b>0$ such that $a h \leq g \leq b h$.

Thus, we have
\[
P_e(\Phi) \geq \frac{1}{8}\exp \pth{ \log \ell - n (c\theta_d^2) },
\]
where $c$ is the constant scaling based on Pinsker's and reverse Pinsker's inequalities.

From this we observe that $N^*_{\text{temp}} = \Omega\pth{\frac{\log \ell - \log \epsilon}{\theta_d^2}}$.
\end{IEEEproof}

\begin{corollary}
Let $f(\cdot)$ be a convex function satisfying the smoothness constraints. Further, let $\min_{i,j \in [\tau]} D_f(Q_i \| Q_j) = \theta_d$. Then for constant $\ell$,
\[
N^*_{\text{temp}} = \Omega(\theta_d^{-1}).
\]
\end{corollary}
\begin{IEEEproof}
From (\ref{eqn:f-div_tvd_bound}) we have 
\[
\min_{i,j \in [\tau]} \delta(Q_i,Q_j) \lesssim \sqrt{\theta_d}.
\]
This proves the result.
\end{IEEEproof}

Thus, we see that the universal clustering decoder achieves the lower bound up to the constant factor in sample complexity. Hence our clustering algorithm is asymptotically order optimal in the number of objects to be clustered and the minimum separation of hypotheses. 

It is worth noting that the quantity $\theta_d^2$ is equivalent in definition to the crowd quality defined in \cite{KargerOS2014} and matches the lower bound obtained on the cost for binary classification using crowd workers biased toward giving the right label. The factor of $\log \ell$ in the cost per object arises since error probability studied here is the block (blocklength $\ell$) error probability whereas\cite{KargerOS2014} studies the average symbol (classification) error probability.

\section{Workers with Memory}

\begin{figure}[t]
	\centering
	\includegraphics[width=4in]{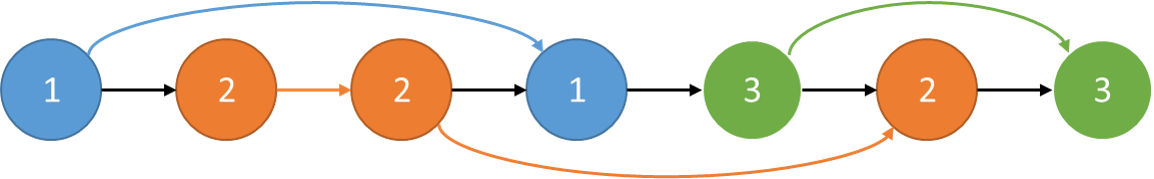}
 	\caption{Bayesian Network model of responses to a set of $7$ objects chosen from a set of $3$ types. We observe that the most recent response and the response to the most recent object of the same type influence every response.}
  	\label{fig:bayes_net}
\end{figure}

Recall that in Section \ref{sec:model}, we defined the structure of the stochastic kernel that determines the responses of workers with memory. In particular, we considered a Markov memory structure \eqref{eqn:memory_kernel}. This structure is represented in the Bayesian network depicted in Fig.~\ref{fig:bayes_net}.

Specifically, we assume that the response $Y_{i,j}$ to an object $X_i$ by worker $j$ is dependent on the response to the most recent object, $X_{i-1}$, and the response to the most recent object of the same class. This set of indices for any object $X_i$ is given by $\calN_i$. 

\subsection{Task Difficulty for Worker Pool}
Let $\mathcal{Q}$ be the set of such Markov-structured distributions representing the worker pool. We assume the workers are chosen independently and identically from this set according to some underlying distribution. Thus, the conditional distribution of the response vector for a random worker is
\begin{align*}
&\prob {Y^\ell=y^\ell \middle\vert T^\ell=t^\ell}  = \expect{Q \pth{ Y^\ell=y^\ell \middle\vert T^\ell=t^\ell}} = \bar{Q}\pth{ y^\ell \middle\vert t^\ell},
\end{align*}
where the expectation is taken over the worker distribution.

We know that a sufficient statistic of the worker responses is the empirical distribution. Asymptotically, we know that the empirical pmf converges to $\bar{Q}$ by the strong law of large numbers. It thus suffices to study the decoder with regard to this characteristic worker response pmf that retains the assumed memory properties. 

Throughout the section, for any $i\in[\ell]$, denote by $\tilde{\imath}$ the index such that $\tilde{\imath} \in \calN_i, T_i = T_{\tilde{\imath}}$. 
\begin{defn}[Memory Quality] \label{def:mem_work}
\emph{The memory quality} in a given pool of long-term workers is
\begin{equation} \label{eqn:mem_crowd_quality}
\theta_m = \frac{1}{2} \pth{\min_{i \in [\ell]} I(Y_i;Y_{\calN_i}) - \max_{i \in [\ell], j < i, j \notin \calN_i} I(Y_i;Y_{i-1},Y_j)}.
\end{equation}
\end{defn}
That is, the memory quality is defined by the difference between the information provided by the neighbors of the object and that provided by two other objects.

Since the crowd responses are defined by conditional independence across objects, we quantify the quality of the crowd in terms of the amount of memory retained by the workers. The above definition can be equivalently viewed as the \emph{task difficulty} for a given pool of long-term crowd workers. 

This informational definition finds operational significance in the coding theorems, Theorems \ref{thm:mem_consistency} and \ref{thm:conv2}.

\subsection{Information Clustering using Neighbors}

From the model of worker responses, we know that identifying the parents of each node is critical to cluster.
\begin{lemma}
Let $G=(V,E)$ be the Bayesian network representation of the worker responses. Let $\nu_i = \{j \in V: (j,i) \in E\}$. If $|\nu_i| = 1$, then either $T_{i-1} = T_i$ or $T_j \neq T_i$ for all $j < i$. If $|\nu_i| > 1$, $\nu_i \backslash \{i-1\}$ is in the same cluster as $i$.
\end{lemma}
\begin{IEEEproof}
The results follow from the model definition.
\end{IEEEproof}

We use the data processing inequality to obtain the following property that motivates the decoder construction.
\begin{lemma} \label{lemma:info_property}
Let $C = \{c_1,\dots,c_k\}$ and without loss of generality, let $c_1<c_2<\dots<c_k$. Given (\ref{eqn:memory_kernel}), $C \in P^*$ if and only if for all $j \in [k]$,
\[
c_j = \arg\max_{i < c_{j+1}} I(Y_{c_{j+1}};Y_{i},Y_{c_{j+1}-1}).
\]
\end{lemma}
\begin{IEEEproof}
Let $i,j \in [\ell]$ such that $j < i$. Then, the result follows from the data processing inequality:
\begin{align*}
I(Y_i;Y_{i-1}, Y_{\tilde{\imath}}, Y_j) &= I(Y_i;Y_{\calN_i}) + I(Y_i; Y_j \vert Y_{\calN_i}) \\
&= I(Y_i;Y_{i-1}, Y_j) + I(Y_i; Y_{\tilde{\imath}} \vert Y_{i-1}, Y_j).
\end{align*}
For the given model, $I(Y_i; Y_j \vert Y_{\calN_i}) = 0$. In the non-trivial problem it is natural that $I(Y_i; Y_{\tilde{\imath}} \vert Y_{i-1}, Y_j) > 0$. That is, the most recent object of the same class has residual information, given any other pair from the past. This in turn implies that for all $j < i, j \neq \tilde{\imath}$,
\[
I(Y_i;Y_{i-1}, Y_j) < I(Y_i;Y_{\calN_i}).
\]
The result thus follows.
\end{IEEEproof}

It is evident that the partition can be obtained through a careful elimination process using mutual information values. Maximum likelihood estimates of the mutual information can be obtained from the samples using and asymptotically consistent estimators. Note that such estimators converge exponentially; convergence rates are detailed in Appendix \ref{app:info_conv}.

\subsection{Information Clustering Algorithm}

We now describe the clustering algorithm in two stages. First we describe an algorithm that, given the set of objects and mutual information values, outputs a partition that is denser than the correct partition. We then describe an algorithm that overcomes this shortcoming by identifying sub-clusters within the identified clusters recursively. We then show correctness of the algorithm and prove it is asymptotically consistent when the ML estimates of mutual information are used.

From the directed acyclic graph (Bayesian network) corresponding to the given set of objects, we know that identifying the parents of each node is sufficient to identify clusters such that objects of the same type are in the same cluster. From Lemma \ref{lemma:info_property}, for any $i \in [\ell]$, $I(Y_i;Y_{i-1}, Y_j) < I(Y_i;Y_{\calN_i})$. Thus identifying the parents of node $i$ is equivalent to solving
\[
\eta_i = \arg\max_{j \leq i-1} I(Y_i;Y_{i-1},Y_j).
\]
Using this feature we design Algorithm \ref{algo:cluster_memory}, $\Phi_{\text{info}}(I)$.

\begin{algorithm}[t]
  \caption{Clustering given MI, $(P) = \Phi_{\text{info}}(I)$}
  \label{algo:cluster_memory}
  \begin{algorithmic}[t]
   \STATE {$F(i) \leftarrow 0$ for all $i \in [m]$}
   \STATE {$P \leftarrow \emptyset$}
   \STATE {$M_c \leftarrow 0$}
   \STATE {$\gamma_n = c_1 n^{-\alpha},$ where $c_1$ is a constant and $\alpha \in (0,1/2)$}
   \FOR{$i=m$ \TO $1$}
    \IF {$F(i) = 0$}
    	\STATE {$F(i) \rightarrow 1$}
    	\STATE {$M_c \leftarrow M_c + 1$, $C(i) \leftarrow M_c$}
    \ENDIF
    \STATE {$I_{\text{max},i} \leftarrow \max_{k \leq i-1} I(Y_i;Y_{i-1},Y_k)$}
    \STATE {$\eta_i = \max\{j < i : I(Y_i;Y_{i-1},Y_j) \geq I_{\text{max},i} - \gamma_n\}$}
    \IF {$F(\eta_i) = 0$}
    	\STATE {$C(\eta_i) = C(i), F(\eta_i) = 1$}
    \ENDIF
   \ENDFOR
   \STATE {$P = \{\{i: C(i) = k\}: k \in [M_c]\}$} 
  \end{algorithmic}
 \end{algorithm}
 
The algorithm outputs the partition of a set of objects when the corresponding mutual information values are given as input. The algorithm starts from the last object and iterates backward while finding the parents of each node. Upon identification, the parent is added to the same cluster as the object. 

\begin{theorem} \label{thm:algo_correctness}
Given a set of objects $T^{\ell}$ and the corresponding set of mutual informations $\{ I(Y_i;Y_{i-1},Y_j) : i \in [\ell], j < i \}$, the output of Alg. \ref{algo:cluster_memory} satisfies $P = \Phi_{\text{info}}(I) \succeq P^*$.
\end{theorem}
\begin{IEEEproof}
From Lemma \ref{lemma:info_property}, $I(Y_i;Y_{i-1},Y_j) \leq I(Y_i; Y_{\calN_i})$ for all $j < i$ with equality if and only if $j \in \calN_i$. Thus, the parents of every node $i \in [\ell]$ in the Bayesian network can be determined, given the mutual information values. 

For $P = \Phi_{\text{info}}(I)$, for every object $t \in \calT$, there exists $C \in P$ such that $\{i \in [\ell]:T_i = t\} \subseteq C$. Hence the result follows.
\end{IEEEproof}

\subsection{Consistency of Universal Clustering}

\begin{figure}
    \centering
            \includegraphics[width=4in]{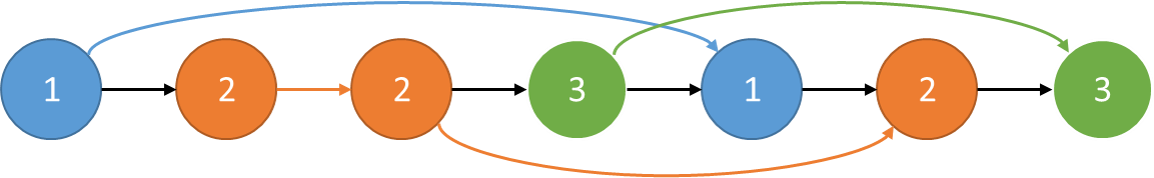} \\ (a) \\
		\includegraphics[width=4in]{bayesian_network.png} \\ (b)

    \caption{Functioning and shortcoming of Alg.~\ref{algo:cluster_memory}.  (a) Alg.~\ref{algo:cluster_memory} outputs $P = \{\{1,5\},\{2,3,6\},\{4,7\}\} =P^*$. Here, object $4$ of type $3$ is not clustered with object $3$ of type $2$ as it is already assigned a cluster. (b) Alg.~\ref{algo:cluster_memory} outputs $P = \{\{1,4,5,7\},\{2,3,6\}\} \succeq P^*$. Here object $5$ is clustered with $4$ as objects of Type $1$ are not encountered before.}
\label{fig:bayes_phi_info}
\end{figure}

Theorem \ref{thm:algo_correctness} indicates that, given the mutual information values, the objects of the same type are clustered together. The maximizer $\eta_i$ in Alg. \ref{algo:cluster_memory} is clustered only if it is not assigned a cluster before the iteration. Thus object $i$ is not paired with $i-1$ unless it has not been assigned a cluster. This particular scenario is depicted in the Bayesian network in Fig.~\ref{fig:bayes_phi_info}(a).

However, the algorithm fails in a specific scenario. When there exists clusters $C_1$ and $C_2$ such that $i<j$ for every $i \in C_1 , j \in C_2$, and, $\max\{i \in C_1\} = \min\{j \in C_2\} - 1$, then the resulting partition consists of the single cluster $C_1 \cup C_2$ rather than the two individual clusters. This is because objects of $C_1$ have not yet been encountered and thus the immediate neighbor of the first object of $C_2$ is clustered along with $C_2$ due to the Markov memory structure. This particular scenario is depicted in Fig.~\ref{fig:bayes_phi_info}(b).

Such shortcomings of the algorithm however happen for outlier cases that are of low probability when $\ell \gg \tau$. Thus, if the finest partition in a collection of permutations of a given set of objects, chosen uniformly at random, is obtained using $\Phi_{\text{info}}$, then with high probability, the correct partition is obtained. Thus the overall algorithm can be summarized as in Alg. \ref{algo:memory_script} when we want an error probability less than $2\epsilon$.

\begin{algorithm}[t]
  \caption{Clustering with memory, $P = \Phi_{\text{mem}}(T^{\ell})$}
  \label{algo:memory_script}
  \begin{algorithmic}[t]
   \STATE {Choose constant $k = \lceil \frac{-\log \epsilon}{(\log \ell - 2\log \tau)} \rceil$}
   \FOR{$i=1$ \TO $k$}
    \STATE {Choose a permutation $\xi([\ell])$ uniformly at random}
    \STATE {Collect responses $\bfY^{(i)}$ for the sequence of objects $T^{\xi([\ell])}$}
    \STATE {Compute $I = \{\hat{I}(Y_j; Y_{j-1},Y_k): k < j, j \in [\ell]\}$}
    \STATE {$P_i = \Phi_{\text{info}}(I)$}
   \ENDFOR
   \STATE {Choose finest partition $P$ such that $P \preceq P_i$ for all $i\in[k]$} 
  \end{algorithmic}
 \end{algorithm}

For each permutation of the objects, $n$ responses are obtained for each object from the workers. Thus, the overall number of samples per object obtained is $kn$.

\begin{theorem} \label{thm:mem_consistency}
Let $T^{\ell}$ be the set of objects, $\ell \geq \tau^2$, and, let $\bfY$ be the set of responses. Let $k \geq \lceil \frac{-\log \epsilon}{(\log \ell - 2\log \tau)} \rceil$ be the number of permutations chosen in Alg. \ref{algo:memory_script}. Then, for 
\begin{equation} \label{eqn:memory_cost_suff}
n \gtrsim \max\sth{\pth{\log \ell - \log \epsilon}^{\frac{1}{(1-2\alpha-\beta)}}, \pth{\log \ell - \log \epsilon}^{\frac{1}{(1-4\alpha)}},\theta_m^{\frac{-1}{\alpha}} },
\end{equation}
for $0 < \alpha < 1/2$ and $0 < \beta < 1$ such that $\log_2^2 n \leq n^\beta$, $P_e(\Phi_{\text{mem}}) \leq 2\epsilon$, for any $\epsilon>0$. Further, for constant $\ell$ and $\theta_m$, the algorithm is asymptotically consistent.
\end{theorem}
\begin{IEEEproof}
We first observe that when $|\hat{I}(Y_i;Y_{i-1},Y_j) - I(Y_i;Y_{i-1},Y_j)| < \theta_m$ for all $i \in [\ell], j<i$, $\Phi_{\text{info}}(\hat{I}) = \Phi_{\text{info}}(I)$. That is, when the empirical mutual information values do not deviate from the actual values significantly, the clustering algorithm works without error.

For $n \geq (c_1/\theta_m)^{1/\alpha}$, $\gamma_n < \theta_m$. Let $I_{ij} = I(Y_i;Y_{i-1},Y_j)$. Then,
\begin{align}
\prob{\Phi_{\text{info}}(\hat{I}) \neq \Phi_{\text{info}}(I)} &\leq \sum_{i\in \ell, j<i}\prob{|\hat{I}_{ij} - I_{ij}| > \gamma_n} \label{eqn:union_bd}\\
&\quad\leq \binom{\ell}{2}\exp\pth{\frac{-n\gamma_n^2}{18 \log_2^2 n} + o(1)} \label{eqn:lem_conc_ref} \\
&\quad\leq \exp\pth{2\log \ell - \frac{c_1^2}{18} n^{(1 - \nu)} + o(1)}, \notag
\end{align}
implying asymptotic consistency. Here $\nu = 2\alpha + \beta$, \eqref{eqn:union_bd} follows from the union bound, and \eqref{eqn:lem_conc_ref} follows from Lemma \ref{lemma:ML_concentration}.

Additionally, we note that
\begin{align}
\prob{\Phi_{\text{info}}(\hat{I}) \neq \Phi_{\text{info}}(I)} &\leq \sum_{i\in \ell, j<i}\prob{|\hat{I}_{ij} - I_{ij}| > \gamma_n} \notag \\
&\quad\leq 3\binom{\ell}{2}\pth{n+1}^{\tau^2}\exp\pth{-\tilde{c}n \gamma_n^4} \label{eqn:second_conv} \\
&\quad\leq \exp\pth{2\log \ell - \tilde{c}c_1^4 n^{(1-4\alpha)} +o(1)}, \notag
\end{align}
where \eqref{eqn:second_conv} follows from \eqref{eqn:inf_cont}.

To obtain the correct partition, we use the responses generated for several uniformly random permutations of the given set of objects and select the finest partition. The correct partition may not be recovered when there exists $t,t' \in \calT$, such that $\max\{i \in [\ell]: T_i = t\} = \min\{i \in [\ell]: T_i = t'\} - 1$ in every chosen partition. 

Let $K_t = |\{i \in [\ell]: T_i = t\}|$, $t \in [\tau]$ and let $M_t = \max \{i \in [\ell]: T_i = t\}$ and $m_t = \min \{i \in [\ell]: T_i = t\}$. Thus, the probability that $P = \Phi_{\text{mem}}(T^{\ell}) \succ P^*$ is bounded as:
\begin{align*}
\prob{P \succ P^*} &= \prob{\exists t \neq t' : M_t = m_{t'} - 1}^k \\
&\leq \pth{ \sum_{t,t' \in \calT, t \neq t'} \prob{M_t = m_{t'} - 1} }^k.
\end{align*} 

The total number of possible permutations of the given set of objects is given by
\[
\kappa_{\text{permut}} = \frac{\ell!}{\prod_{\tilde{t} \in \calT} K_{\tilde{t}}!}.
\] 
The number of sequences such that $M_t = m_{t'} - 1$ can be determined by choosing $K_t + K_{t'} - 1$ locations out of $\ell - 1$ locations to fill the objects of type $t$ and $t'$ and permute over the other objects. Thus,
\[
\kappa_{\text{t,t'}} \leq \binom{\ell - 1}{K_t + K_{t'} - 1} \frac{(\ell - K_t - K_{t'})!}{\prod_{\tilde{t} \in \calT\backslash\{t,t'\}} K_{\tilde{t}}!}.
\]
Since the permutations are chosen uniformly at random,
\begin{align}
\prob{ m_{t'} - M_t = 1} &\leq \frac{\kappa_{t,t'}}{\kappa_{\text{permut}}} \notag \\
&= \frac{1}{\ell} \frac{K_t! K_{t'}!}{(K_t + K_{t'} - 1)!} \leq \frac{1}{\ell}, \label{eqn:permut_ineq}
\end{align}
where (\ref{eqn:permut_ineq}) follows from the fact that
\begin{align}
\frac{K_t!K_{t'}!}{(K_t+K_{t'}-1)!} &\leq 1,
\end{align}
as $K_t, K_{t'} \geq 1$.

Thus, we have
\begin{align}
\prob{P \succ P^*} &\leq \pth{\binom{\tau}{2}\frac{2}{\ell}}^k \leq \exp\pth{-k (\log \ell - 2\log \tau)}.
\end{align}
Thus, for $k \geq \lceil \frac{-\log \epsilon}{(\log \ell - 2\log \tau)} \rceil$, $\prob{P \succ P^*} \leq \epsilon$.

Now we prove consistency of Alg. \ref{algo:memory_script}. Using the union bound, we observe that the probability of error is bounded as
\begin{align}
P_e &\leq \prob{P \succ P^*} + k\prob{\Phi_{\text{info}}(\hat{I}) \neq \Phi_{\text{info}}(I)} \notag \\
&\leq \exp\pth{-k (\log \ell - 2\log \tau + \log 2)} \notag \\
&\quad + \exp\pth{2\log \ell - \max\sth{\frac{c_1^2}{18} n^{(1-\nu)}, \tilde{c}c_1^4 n^{(1-4\alpha)}}  + o(1)} \label{eqn:two_conc_results} \\ 
&\leq 2\epsilon \notag
\end{align}
for a large enough $n$. Here (\ref{eqn:two_conc_results}) follows from the two concentration bounds on empirical mutual information described in Appendix \ref{app:info_conv}. 

Thus, for any $\epsilon > 0$, there exists $n,k$ sufficiently large, such that $P_e < \epsilon$. Hence the consistency result follows. The sample complexity is obtained from the error exponent in (\ref{eqn:two_conc_results}).
\end{IEEEproof}

We observe from the proof that there is a trade-off between the values of $k$ and $n$ needed to achieve a certain level of accuracy. In particular, we observe that when $\ell$ is large, it suffices to consider a small number of permutations of the set of objects, while each permutation requires a larger number of samples. On the other hand, when $\ell$ is relatively small, one needs a large number of permutations while each permutation requires far fewer samples. 

We restrict focus to the case where $\ell \gg \tau^2$ and under this scenario find the following result on sample complexity.
\begin{corollary} \label{cor:sample_comp_UB}
Given $\calT = [\tau]$ with $\tau < \infty$ a constant and $\ell \gg \tau^2$,
\[
N^*_{\text{mem}}(\epsilon) = O\pth{\frac{\pth{\log \ell - \log \epsilon}^{\min\sth{1/(1-2\alpha-\beta), 1/(1-4\alpha)}}}{{\theta_m^{(2/(1-\beta))}} }}.
\]
\end{corollary}
\begin{IEEEproof}
Using Theorem \ref{thm:mem_consistency} and the fact that the total number of samples used is $kn$ (since clustering with $n$ samples is done $k$ times) per object, we obtain the result. 
\end{IEEEproof}
For large $\ell$, we can thus observe that $N^*_{\text{mem}}(\epsilon) = O\pth{\frac{\log \ell}{\theta_m^2}}$.

Note that under the Markov memory model for long-time workers, the sufficient number of samples per object is the same in order as for temporary workers.

\subsection{Lower Bound on Sample Complexity}

We now provide matching lower bounds by studying the probability of error of a problem which is a reduction of the universal clustering problem.

\begin{theorem} \label{thm:conv2}
The sample complexity of universal clustering using workers with memory satisfies
\begin{enumerate}
\item for a fixed $\theta_m > 0$, $N^*_{\text{mem}} = \Omega \pth{\log \ell}$, and
\item for a fixed $\ell < \infty$, $N^*_{\text{mem}} = \Omega \pth{\theta_m^{-1}}$.
\end{enumerate}
\end{theorem}
\begin{IEEEproof}
Choose a prior, parametrized by the size of the problem $\ell$ as $P_T(1) = 1-\frac{1}{\ell}, P_T(2) = \frac{1}{\ell}$ such that $\frac{1}{\ell} = \frac{1}{\ell}$. Let $\calE$ be the set of all vectors of objects with at most one object of type $2$. Then, 
\[
\prob{T^\ell \in \calE} = \pth{2-\frac{1}{\ell}}\pth{1-\frac{1}{\ell}}^{\ell-1} \stackrel{\ell \rightarrow \infty}{\longrightarrow} 1.
\] 
In particular, we note that $\prob{T^\ell \in \calE}$ is an increasing function of $\ell$ and is at least $\tfrac{1}{2}$ for any $\ell > 3$.

For a given constant $\theta_m$, consider the special case of the problem where $\calN_i = \{\tilde{\imath}\}$. That is, consider the problem where any two objects are dependent if and only if they are of the same type. Clearly, any algorithm that solves the universal clustering with memory problem solves this simplified problem as well. Thus, following the convention established, we have
\[
\begin{cases}
I(Y_i;Y_{\tilde{\imath}}) \geq 2\theta_m, & \text{ for all } i \in [\ell] \\
I(Y_i;Y_j) = 0, & \text{ for all } i,j \in [\ell], T_i \neq T_j
\end{cases}.
\]

Define 
\[
W = [W_{ij}]_{1 \leq i,j \leq 2} = \begin{bmatrix}
\frac{1}{2} + \epsilon & \frac{1}{2} - \epsilon \\
\frac{1}{2} - \epsilon & \frac{1}{2} + \epsilon
\end{bmatrix}.
\]
Consider the scenario where worker responses are inertial over time and characterized as:
\[
\prob{Y_i = k \vert Y_{\tilde{\imath}} = j} = W_{kj},
\]
for any $k,j \in \{0,1\}$ and $i \in [\ell]$. Additionally, assume that the marginals of the responses are uniform (that is, the response to the first object of each type is distributed as Bern$(1/2)$). 

The information constraint implies
\[
\frac{1}{2} - h^{-1}(1-2\theta_m) \leq \epsilon < \frac{1}{2},
\]
where $h(\cdot)$ is the binary entropy function and $h^{-1}(\cdot)$ is its inverse. Let $\epsilon = \frac{1}{2} - h^{-1}(1-2\theta_m)$.

From the definition of the error probability,
\begin{align}
P_e &\geq \frac{1}{2} \prob{\hat{P} \neq P^* \vert T^{\ell} \in \calE}.
\end{align}
Now consider the set $\calE$ of vectors. Identifying the correct partition for a vector of objects from this space is equivalent to identifying the objects. Thus consider the $(\ell+1)$-ary hypothesis testing problem defined by
\begin{equation} \label{eqn:m-ary}
\begin{cases} 
H_0 : T_j = 1, \text{ for all } j \in [\ell] \\
H_i : T_i = 2, T_j = 1, \text{ for all } j \neq i
\end{cases}.
\end{equation}

We seek to compute the average error probability of \eqref{eqn:m-ary} corresponding to the prior $P_T$. Due to symmetry, note that the optimal decoder accrues the same probability of error under $H_i$ for any $i>0$. Thus
\begin{align*}
\prob{\text{error in (\ref{eqn:m-ary})}} &= \prob{H_0}\prob{\text{error in (\ref{eqn:m-ary})} \vert H_0} \\
&\quad + \prob{\text{error in (\ref{eqn:m-ary})} \vert H_1} \pth{\sum_{i \in [\ell]} \prob{H_i}}.
\end{align*}

Now, note that
\[
\sum_{i \in [\ell]} \prob{H_i} = \pth{1-\frac{1}{\ell}}^{\ell - 1}, \quad \prob{H_0} = \pth{1-\frac{1}{\ell}}^{\ell}.
\]
Thus, for $\ell > 1$,
\[
\frac{1}{2} \prob{\{H_i: i>0\}} \leq \prob{H_0} \leq \prob{\{H_i: i>0\}}.
\]
Thus, $\prob{\{H_i: i>0\}} \asymp \prob{H_0}$. This indicates that the average error probability is lower-bounded by a constant factor of the minimax error probability lower bound for (\ref{eqn:m-ary}). 

Let $Q_i$ be the distribution of the set of responses corresponding to the hypotheses defined in (\ref{eqn:m-ary}):
\begin{equation}
Q_i(Y^\ell) = \begin{cases}
\frac{1}{2} \prod_{k=2}^\ell W_{Y_k Y_{k-1}}, &  i=0 \\ 
\frac{1}{4} \prod_{j=2}^{i-1} W_{Y_j  Y_{j-1}} \prod_{k=i+1}^\ell W_{Y_k Y_{k-1}}, &  i\in[\ell]
\end{cases}.
\end{equation}

\begin{lemma} \label{lemma:hypotheses_KL_bound}
For all $\ell$, $D\pth{Q_i \| Q_j} = O(1)$.
\end{lemma}
\begin{IEEEproof}
See Appendix \ref{app:KL_bound_proof}.
\end{IEEEproof}
Having bounded the KL divergences between the hypotheses, we obtain a lower bound on the error probability of \eqref{eqn:m-ary} using the generalized Fano inequality \cite{Yu1997}. 

Let $\beta = \max_{i,j \in [\ell]\cup\{0\}, i \neq j} D(Q_i \| Q_j)$. The loss function considered here is the 0-1 loss. Hence,
\begin{align}
\prob{\hat{P} \neq P^* \vert T^{\ell} \in \calE} &= \prob{\text{error in problem (\ref{eqn:m-ary})}} \notag \\
&\asymp \max_{0 \leq i \leq \ell} \prob{\text{error in problem (\ref{eqn:m-ary})} \vert H_i} \notag \\
&\geq \frac{1}{2}\pth{1 - \frac{n\beta + \log 2}{\log (\ell+1)}}. \label{eqn:error_Fano_ineq}
\end{align} 

Hence, for a constant $\theta_m > 0$, the sample complexity of universal clustering satisfies
\[
N^*_{\text{mem}} = \Omega(\log \ell).
\]

Now, when $\ell$ is fixed, we seek to understand the sample complexity with respect to the memory quality of the crowd. To this end, we note that any consistent clustering algorithm is also consistent for the binary hypothesis test
\begin{equation} \label{eqn:binary_mem_ij}
\psi:\begin{cases}
H_0: I(Y_1;Y_2) = 0  \\
H_1: I(Y_1;Y_2) \geq 2\theta_m
\end{cases}.
\end{equation}
That is, if $\Phi$ is a decoder for the universal clustering problem, then $\Phi$ also solves $\psi$.

Since the sufficient statistics for detection of the binary hypothesis testing above are the responses to $T_1,T_2$, it suffices to consider $Y_1^n,Y_2^n$. Let the prior here be $P_T(1) = P_T(2) = 1/2$. Let the corresponding distributions of worker responses be $p(Y_1^n,Y_2^n)$ and $q(Y_1^n,Y_2^n)$ under $H_0$ and $H_1$ respectively. Here,
\begin{align*}
p(y_1^n,y_2^n) &= \frac{1}{2} \prod_{i=1}^n \prob{y_{1,i},y_{2,i} \vert T_1 = T_2 = 1} \\
& \quad + \frac{1}{2} \prod_{i=1}^n \prob{Y_{1,i},Y_{2,i} \vert T_1 = T_2 = 2},
\end{align*}
and
\begin{align*}
q(Y_1^n,Y_2^n) &= \frac{1}{2} \prod_{i=1}^n \prob{Y_{1,i} \vert T_1 = 1}\prob{Y_{2,i} \vert T_1 = 2} \\
& \quad + \frac{1}{2} \prod_{i=1}^n \prob{Y_{1,i} \vert T_1 = 2}\prob{Y_{2,i} \vert T_1 = 1}
\end{align*}

Let $p_j^{(i)}(y) = \prob{Y_i = y \vert T_i = j}$, $q_j^{(i)}(y) = \prob{Y_i = y \vert T_i = j}$,
under $H_0$ and $H_1$ respectively. Without loss of generality, we assume $I_p(Y_1,Y_2) = 2\theta_m$.

Since the distributions satisfy the information constraints, we have,
\begin{align}
D(p \| q) &\leq  \frac{n}{4}\pth{4 I_{p}( Y_1;Y_2 ) + \sum_{i,j,k \in [2]} D( p_j^{(i)} \| q_k^{(i)} ) } \label{eqn:Jensen_to_info} \\
&= 2n\theta_m, \notag
\end{align}
when the marginals under the two hypotheses are equal as was the case in the inertial worker response channel. Here (\ref{eqn:Jensen_to_info}) follows from convexity. Thus the minimum upper bound on the KL divergence between the hypotheses is $2\theta_m$. Since we consider the worst case with respect to $\theta_m$, it suffices to consider this upper bound.

Thus,
\begin{align}
P_e(\Phi) &\geq \prob{\text{error in }\psi} \geq \frac{1}{4}\exp\pth{-2B(p,q)} \label{eqn:Kailath_LB_2} \\
&\geq \frac{1}{4}\exp\pth{-2D(p \| q )} \geq \frac{1}{4}\exp\pth{-4n\theta_m}, \label{eqn:Jensen}
\end{align}
where (\ref{eqn:Kailath_LB_2}) follows from the Kailath lower bound \cite{Kailath1967}. Then, using Jensen's inequality, we obtain \eqref{eqn:Jensen}. Thus 
\[
N^*_{\text{mem}} = \Omega\pth{\theta_m^{-1}}.
\]
\end{IEEEproof}

We observe from Theorem~\ref{thm:conv2} that the universal clustering algorithm is order optimal in terms of the number of objects, $\ell$. However, there is a gap between the lower bound and the achievable cost in terms of the crowd quality $\theta_m$. This gap is exactly the well-known gap for entropy estimation observed in \cite[Corollary 2]{ValiantV2011}.

\subsection{Reductions to Other Clustering Algorithms}

There exist several clustering paradigms based on mutual information. Here we describe two such algorithms and reductions of our model that lead to our clustering algorithm becoming the same as those algorithms.

An information clustering strategy is defined in \cite{ChanAEKL2015} that identifies clusters based on the minimum partition information that is defined as
\[
I(Z_V) = \min_{P \in \calP} I_P(Z_V),
\]
where $I_P$ is the partition information according to the partition $P$, defined as
\[
I_P(Z_V) = \frac{1}{|P|-1} \pth{\sum_{C \in P} H(Z_C) - H(Z_V)}.
\]

Consider the Markov memory model such that $\calN_i = \{\tilde{\imath}\}$, that is, if $Y_i$ is conditionally dependent on an object only if it is of the same type. Then, if $|P^*| > 1$, then
\[
I(Y^{\ell}) = I_{P^*} (Y^{\ell}) = 0.
\]
That is, the correct partition is the partition that minimizes the partition information. The following reduction indicates that minimizing the partition information is the same as our algorithm, and so \cite{ChanAEKL2015} is a special case of our approach. 

First, we have
\begin{align*}
H(Y^\ell) &= \sum_{i=1}^\ell H(Y_i \vert Y_{\tilde{\imath}}).
\end{align*}
Similarly, 
\[
H(Y_C) = \sum_{i \in C} H(Y_{i} \vert Y_{j(i)}),
\]
where $j(i) = \max\{k < i : k \in C\}$. This implies that
\begin{align*}
I_P(Y^\ell) &= \frac{1}{|P|-1} \pth{\sum_{i=1}^\ell (I(Y_i; Y_{\tilde{\imath}}) - I(Y_i; Y_{j(i)}))},
\end{align*}
where $j(i) = \max\{k < i : i,k \in C\}$. This indicates that minimizing the partition information is equivalent to $\Phi_{\text{info}}(I)$.

Another information-based clustering strategy is the mutual information relevance network (MIRN) clustering \cite{NaganoKI2010}. Here, for a given threshold $\gamma$, the clustering strategy determines the connected components of $G = ([\ell], E)$ such that $(i,j) \in E \iff I(Z_i;Z_j) > \gamma$. For the Markov memory model, under the restriction that 
\[
\min_{\{i,j \in [\ell]: T_i = T_j\}} I(Y_i; Y_j) > \gamma \geq \max_{\{i,j \in [\ell]: T_i \neq T_j\}} I(Y_i; Y_j),
\]
MIRN outputs the correct partition. 

However, in the universal clustering scenario, the decoder is not aware of $\gamma$ and thus, it may not be feasible to implement MIRN clustering optimally. Nevertheless. under the restriction that $\calN_i = \{\tilde{\imath}\}$ and $\gamma = 0$, the MIRN clustering algorithm is just a special case of $\Phi_{\text{info}}(I)$.

\subsection{Extended Memory Workers}

While the model defined above considers the dependence of responses on just the most recent object of the same kind, our results hold for any fixed, finite-order Markov structure as well. In particular, consider the scenario where worker responses are dependent on the set $\calN_i = \{i-1\} \cup \calM_i$, where $\calM_i \subseteq \{j < i: T_j = T_i\}$ such that it contains the most recent $\zeta$ indices of the same type of object. That is, the response to an object is dependent not only on the most recent response, but also a constant number of prior responses to objects of the same type.

\begin{figure}[t]
	\centering
	\includegraphics[scale=0.5]{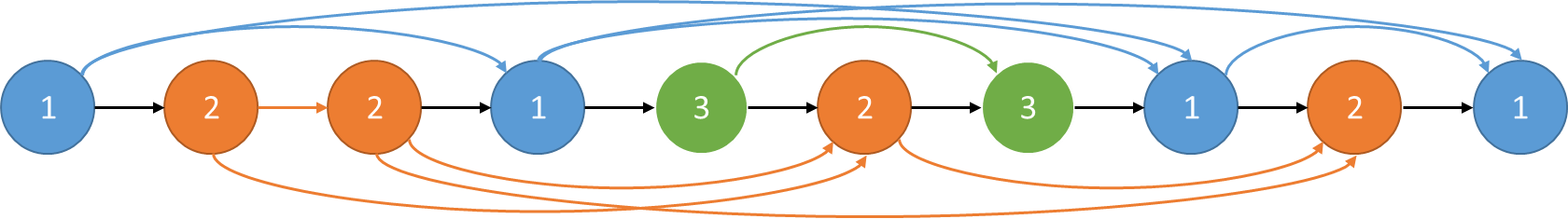}
 	\caption{Bayesian Network model of responses of to a set of $10$ objects chosen from a set of $3$ types with $\zeta = 2$. We observe that the most recent response and the response to the two most recent object of the same type influence every response.}
  	\label{fig:unified_bayesian}
\end{figure}

An example of this worker model can be found in Fig.~\ref{fig:unified_bayesian} for a set of $10$ objects of $3$ types with a worker memory of $\zeta = 2$. 

Then the algorithm defined and the results obtained can be extended to this scenario. More specifically, the parents of a node $i$ can be computed using the rule
\[
I(Y_i;Y_{\calN_i}) > I(Y_i;Y_S),
\]
for any index set $S \subseteq [i-1]$ such that $|S| \leq \zeta$. 

Then for any constant $\zeta$ and $n \gtrsim (\tau+1)^{2\zeta}$, the consistency of the algorithm holds. That is, as long as $\zeta = O(1)$, the sample complexity results follow.

\section {Unified Worker Model}

While we studied two distinct classes of worker models in temporary workers and workers with memory, these two scenarios are limiting cases of a unified worker model described here. After all, it is reasonable to characterize practical crowd worker decisions as influenced by both aspects---memory of individual responses and task difficulty with respect to objects.

For the unified model, we provide an achievable scheme that makes use of the algorithms defined earlier. Further, we prove consistency and order optimality of the scheme.
As in Section \ref{sec:model}, consider worker model \eqref{eqn:unified_kernel}, where each worker is characterized by a Markov memory model subject to fixed conditional marginal distributions.

\subsection {Worker Quality}

Let us now define worker quality. Let $Q_1,\dots,Q_{\tau}$ be the conditional response distributions given the object class, defined as
\[
Q_i(y) \triangleq \prob{Y = y \vert T = i} = \expect{Q^{(W)} ( Y = y \vert T = i)},
\]
where the expectation is taken over the workers in the pool. Define \emph{distance quality} as
\[
\theta_d \triangleq \min_{\{i,j \in \calT, i \neq j\}} \delta(Q_i\|Q_j).
\]
Notice this \emph{distance quality} is analogous to the definition of the worker quality in the case of temporary workers.

Additionally, define the \emph{memory quality} as
\[
\theta_m \triangleq \frac{1}{2} \pth{\min_{i \in [\ell]} I(Y_i;Y_{\calN_i}) - \max_{i \in [\ell], j < i, j \notin \calN_i} I(Y_i;Y_{i-1},Y_j)}.
\]
Notice this \emph{memory quality} is analogous to the definition of worker quality defined for the worker with memory scenario.

Worker quality in the unified worker model is a combination of these individual quality parameters.

\subsection {Clustering Algorithm}

We now provide the universal clustering strategy for the unified worker model, Alg.~\ref{algo:unified_clustering}, drawing on achievable schemes from before.
First perform the memory-based clustering defined in Alg.~\ref{algo:memory_script}; then for every cluster in the partition output by the algorithm, perform distance-based clustering. We now show the consistency of the algorithm.

\begin{algorithm}[t]
  \caption{Clustering under unified worker model, $\Phi_{\text{u}}(T^\ell)$}
  \label{algo:unified_clustering}
  \begin{algorithmic}[t]
   \STATE {$P_{\text{info}} \leftarrow \Phi_{\text{mem}}(T^\ell)$}
   \FOR {$C \in P_{\text{info}}$}
   	\STATE {$P_C \leftarrow \Phi_{\text{temp}}(T^C)$}
   \ENDFOR
   \STATE {$P \leftarrow \cup_{C \in P_{\text{info}}} P_C$}
  \end{algorithmic}
 \end{algorithm}

\begin{theorem}
Let $T^\ell$ be the set of objects and $\ell > \tau^2$. Then, for
\begin{align}
&n \gtrsim \max\left\{\pth{\log \ell - \log \epsilon}^{\frac{1}{(1-2\alpha-\beta)}}, \right. \\
&\qquad\qquad\left.\pth{\log \ell - \log \epsilon}^{\frac{1}{(1-4\alpha)}},(\theta_m + \theta_d)^{\frac{-1}{\alpha}}\right\},
\end{align}
for $0 < \alpha < 1/2$ and $0< \beta < 1$, $P_e(\Phi_{\text{u}}) \leq 2\epsilon$, for any $\epsilon>0$.
\end{theorem}
\begin{IEEEproof}
First we note that for $n \geq (c_1/(\theta_m + \theta_d/4))^{1/\alpha}$, $\gamma_n \leq \theta_d/4 + \theta_m$. Thus, at least one of $\gamma_n \leq \theta_d/4$ or $\gamma_n \leq \theta_m$ is true. This in turn indicates that at least one of $\Phi_{\text{mem}}$ or $\Phi_{\text{temp}}$ is consistent.

Next, from Theorem \ref{thm:algo_correctness}, we note that the output $P_{\text{info}} \succeq P^*$. Thus, subsequent clustering of the individual clusters is sufficient. This in turn indicates the correctness and asymptotic consistency of Alg. \ref{algo:unified_clustering}.
\end{IEEEproof}

We now observe that the sample complexity with respect to the number of objects to be clustered is still $O(\log \ell)$ while that with respect to the quality parameters is $O((\theta_m + \theta_d)^{-2})$.
\begin{corollary}
Given $\calT = [\tau]$ with $\tau < \infty$ a constant,
\begin{enumerate}
\item for a constant $\theta_m, \theta_d > 0$, $N_{\text{u}}^*(\epsilon) = O\pth{\log \ell}$;
\item for a constant $\ell$, $N_{\text{u}}^*(\epsilon) = O((\theta_d + \theta_m)^{-2})$.
\end{enumerate}
\end{corollary}

It is worth noting the limiting cases of the unified worker model. In particular, when $\theta_m \rightarrow 0$, the problem reduces to clustering with temporary workers as do the achievable scheme and sample complexity requirements. On the other hand, $\theta_d \rightarrow 0$ corresponds to a particular case of clustering using workers with memory.

\subsection {Lower Bound on Sample Complexity}

We now derive the lower bound on sample complexity by extending the proof of the converse for workers with memory.
\begin{theorem}
Sample complexity of universal clustering under the unified worker model satisfies
\begin{enumerate}
\item for a fixed $\theta > 0$, $N^*_{u} = \Omega \pth{\log \ell}$, and
\item for a fixed $\ell < \infty$, $N^*_{u} = \Omega \pth{(\theta_m+\theta_d^2)^{-1}}$.
\end{enumerate}
\end{theorem} 
\begin{IEEEproof}
We proceed in similar fashion to the proof for the case of workers with memory. Again, consider the prior parametrized by the size of the problem $\ell$ as $P_T(1) = 1-\pi, P_T(2) = \pi$ such that $\pi = \frac{1}{\ell}$. Again, we will use the generalized Fano's inequality over the space $\calE$ of vectors of objects. We again consider the case of $\calN_i = \{\tilde{i}\}$.

Consider worker responses such that marginals of the responses to an object satisfy
\[
\prob{Y = i \vert T = j} = \begin{cases}
p, & i=j \\
1-p, & i\neq j
\end{cases}
\]
irrespective of the order of occurrence. Define the matrices
\[
W^{(1)} = [W_{ij}^{(1)}]_{1 \leq i,j \leq 2} = \begin{bmatrix}
a & 1-a \\
1-b & b
\end{bmatrix},
\]
and
\[ 
W^{(2)} = [W_{ij}^{(2)}]_{1 \leq i,j \leq 2} = \begin{bmatrix}
b & 1-b \\
1-a & a
\end{bmatrix}.
\]
Let the worker responses be characterized by
\[
\prob{Y_i = k \middle\vert Y_{\tilde{\imath}} = \tilde{k}, T_i = T_{\tilde{\imath}} = j} = W_{k\tilde{k}}^{(j)}.
\]

From the constraint on distance quality, we have:
\[
2p - 1 \geq \theta_d.
\]
The constraint on the nature of the marginals establishes:
\[
ap - b(1-p) = p.
\]
The restriction on the information quality implies:
\[
h(p) - ph(a) - (1-p)h(b) \geq 2\theta_m.
\]
Let us consider the case when both inequalities hold with equality. This yields a specific worker channel that satisfies the memory and distance quality requirements. We analyze the error probability on this worker channel.

Again, using analysis similar to the proof of Lemma \ref{lemma:hypotheses_KL_bound}, we observe the KL divergences between the hypotheses in the $(\ell+1)$-ary hypothesis testing problem are $O(1)$. Hence there exists a constant $\beta$ such that \eqref{eqn:error_Fano_ineq} holds. Thus, for constant $\theta_m$ and $\theta_d$, the sample complexity of universal clustering satisfies:
\[
N^*_{u} = O(\log \ell).
\]

Now, when $\ell$ is fixed, we study the necessary sample complexity of universal clustering with respect to $\theta_m,\theta_d$. We know that a consistent universal clustering algorithm also solves the binary hypothesis test
\[
\psi:\begin{cases}
H_0: I(Y_1;Y_2) = 0  \\
H_1: I(Y_1;Y_2) \geq 2\theta_m.
\end{cases}
\]
Following the analysis from the proof of Theorem \ref{thm:conv2}, from (\ref{eqn:Jensen_to_info}), we have
\[
D(p\|q) \leq n \pth{2\theta_m + \theta_d \log\pth{\frac{1+\theta_d}{1-\theta_d}}} \lesssim n(\theta_m+\theta_d^2).
\]

Finally, using the Kailath lower bound, we obtain
\begin{equation}
P_e(\Phi) \geq \frac{1}{4} \exp\pth{-4cn(\theta_m+\theta_d^2)}.
\end{equation}
Thus, for constant $\ell$,
\[
N^*_u = \Omega((\theta_m+\theta_d^2)^{-1}).
\]
\end{IEEEproof}

From the theorem, we note the universal clustering algorithm is order optimal in sample complexity in terms of the number of objects, for a crowd of given quality. However, for a given number of objects, there exists an order gap between achievable sample complexity and the converse. As expected, the gap follows from the gap in the case of workers with memory, which in turn is from the gap in estimating entropy \cite{ValiantV2011}.

In particular, we observe that in the limit of $\theta_d \rightarrow 0$, the problem reduces to the case of workers with memory and on the other hand the case of $\theta_m \rightarrow 0$ reduces to the problem of clustering using temporary workers without memory. 

A finer point in the analysis to be noted is that the worst-case channels considered in the converse proofs is the inertial channel considered in the proof of Theorem \ref{thm:conv2}, which is also the solution to the set of constraints for the channel in the unified scenario under the limit of $\theta_d \rightarrow 0$.

Thus, we observe temporary workers and long-term workers with memory are indeed closely related through the unified worker model studied here and are limiting scenarios.

\section{Conclusion}

This paper establishes an information-theoretic framework to study the universal crowdsourcing problem. Specifically, we defined a unified worker model (incorporating aspects of human decision making from experimental crowdsourcing and behavioral economics) and designed unsupervised clustering algorithms that are budget optimal. We first studied two limiting cases of workers---ones with and ones without memory.

For temporary workers without memory, we used distributional identicality of responses to design a universal clustering algorithm that is asymptotically consistent and order optimal in sample complexity. For workers with memory, under a Markov model of memory, we used the dependence structure to design a novel universal clustering algorithm that is asymptotically consistent and order optimal in sample complexity with respect to the number of objects. We also note that the gap obtained between necessary and sufficient conditions on sample complexity with respect to the memory quality is also observed in the empirical estimation of entropy.

We then integrated the limiting cases to develop a universal clustering algorithm for the unified worker model. We again proved asymptotic consistency and order optimality in sample complexity with respect to the number of objects. With regard to the quality of crowd workers, the gap observed in the case of the workers with memory remains. 

Behavioral experiments using crowd workers on platforms such as Amazon MTurk can be performed to gain insight into the performance of the algorithms in practice and to validate the unified worker models.

Our results provide a way to compare costs between crowdsourcing platforms having workers with and without memory, thereby providing the opportunity to choose the right task-dependent platform. Further, they provide a window into more general studies of the computational capabilities and complexities of human-based information systems. In particular, the work sheds light on the influence of various attributes of crowd workers such as object-specific memory. In essence, the work studies a space-time tradeoff for human computation systems and to the best of our knowledge is the first of its kind.

\appendices

\section{Concentration of Empirical Distributions} \label{app:dist_conv}
In this section we briefly study the rates of convergence of the ML estimates of $f$-divergence.

Let $\calZ$ be a finite set of objects and $Z_1,\dots,Z_n \stackrel{iid}{\sim} p$. Let $\hat{p}$ be the empirical distribution obtained as 
\begin{equation*}
\hat{p}(z) = \frac{1}{n}\sum_{i=1}^n \indc{Z_i = z}.
\end{equation*}

\begin{lemma} \label{lemma:emp_conc}
If $p$ and $\hat{p}$ are the true and empirical distributions respectively, then
\begin{equation} \label{eqn:emp_conc_dist}
\prob{\delta(\hat{p},p) \ge \epsilon} \le (n+1)^{|\calZ|} \exp \pth{-c_0 n\epsilon^2},
\end{equation}
where $c_0 = 2\log_2 e$. Further, for any convex function $f$ satisfying the smoothness constraints, 
\begin{equation}
\prob{D_f(\hat{p} \| p) \ge \epsilon} \le (n+1)^{|\calZ|} \exp \pth{-n\epsilon/C},
\end{equation}
where $C < \infty $ is a constant such that $xf''(x) < C$.
\end{lemma}
\begin{IEEEproof}
The results follow from Pinsker's inequality (\ref{eqn:Pinsker}), Theorem \ref{thm:f-div_bound}, and Sanov's theorem.
\end{IEEEproof}

\section{Estimating Mutual Information from Samples} \label{app:info_conv}

Here we briefly describe the maximum likelihood (ML) estimate of mutual information and its convergence properties.

Let $X \sim p$ be a random variable on a discrete space $\calX$. Let $X_1,\dots,X_n \stackrel{\text{iid}}{\sim} p$ and let $\hat{p}$ be the corresponding empirical distribution. The ML estimate of entropy of $X$ is given by
\[
\hat{H}(X) = \Expect_{\hat{p}}\qth{-\log (\hat{p}(X))}.
\]
The ML estimate of mutual information between random variables $X,Y$ is then given by
\[
\hat{I}(X;Y) = \hat{H}(X) + \hat{H}(Y) - \hat{H}(X,Y).
\]

The ML estimates of entropy and mutual information have been widely studied \cite{AntosK2001, Paninski2003, ValiantV2011, NetrapalliBSS2010}. In particular, the following results are notable:
\begin{enumerate}
\item  For all $n$, from Jensen's inequality, $\Expect [\hat{H}(X)]<H(X)$ \cite{AntosK2001}. For all $n$, $\epsilon>0$,
\begin{equation}\label{eqn:ent_conv_expectation}
\prob {|\hat{H}(X)- \Expect [\hat{H}(X)]|>\epsilon} \leq 2\exp\pth{ \frac{-n\epsilon^2}{2\log_2^2 n}},
\end{equation}
from McDiarmid's inequality.
\item ML estimate, $\hat{H}$ is negatively biased \cite{Paninski2003}:
\[
b_n(\hat{H}) \triangleq \Expect_p \qth{\hat{H}(X)} - H(X) = -\Expect_p \qth{D(\hat{p} \| p)} < 0.
\]
Further,
\[
-\frac{|\calX|}{n} \leq -\log \pth{1+\frac{|\calX| - 1}{n}} \leq b_n(\hat{H}) \leq 0.
\]
\item From \cite{ValiantV2011}, we know that the lower bound on sample complexity for estimating entropy up to an additive error of $\epsilon$ is $\frac{|\calX|}{\epsilon\log |\calX|}$.
\item From \cite{NetrapalliBSS2010}, we observe that the deviation of the empirical entropy from the entropy of $X \sim P$, is bounded in terms of the total variational distance as
\begin{equation} \label{eqn:ent_cont}
|\hat{H}(X) - H(X)| \leq -2\delta(\hat{P},P) \log \frac{2\delta(P,Q)}{|\calX|}.
\end{equation}
\end{enumerate}

Since we deal with finite, constant alphabet sizes, it suffices for us to consider the ML estimates with sufficiently large $n$, such that the bias is negligible.
\begin{lemma} \label{lemma:ML_concentration}
For fixed alphabet sizes, $|\calX|, |\calY|$, the ML estimate of entropy and mutual information are asymptotically consistent and satisfy
\begin{equation}
\prob{|\hat{H}(X) - H(X)| > \epsilon} \leq 2\exp\pth{\frac{-n\epsilon^2}{2 \log_2^2 n} + o(1)}
\end{equation}
\begin{equation}
\prob{|\hat{I}(X;Y) - I(X;Y)| > \epsilon} \leq 6\exp\pth{\frac{-n\epsilon^2}{18 \log_2^2 n} + o(1)}.
\end{equation}
\end{lemma}
\begin{IEEEproof}
The convergence of entropy follows directly by applying the triangle inequality, union bound, and \eqref{eqn:ent_conv_expectation}. The result follows from the fact that the alphabet is of finite, constant size. This implies the convergence result for mutual information.
\end{IEEEproof}

\begin{lemma} \label{lemma:ML_conc_2}
For fixed alphabet sizes, $|\calX|,|\calY|$, the ML estimate of entropy and mutual information are asymptotically consistent and satisfy
\begin{equation}
\prob{|\hat{H}(X) - H(X)| > \epsilon} \leq (n+1)^{|\calX|} \exp \pth{-c n \epsilon^4},
\end{equation}
\begin{equation}
\prob{|\hat{I}(X;Y) - I(X;Y)| > \epsilon} \leq 3(n+1)^{|\calX||\calY|} \exp \pth{-\tilde{c} n \epsilon^4}, \label{eqn:inf_cont}
\end{equation}
where $c = \pth{2|\calX|^2\log2}^{-1}$, $\tilde{c} = \pth{32\max\{|\calX|,|\calY|\}^2\log2}^{-1}$.
\end{lemma}
\begin{IEEEproof}
We first observe that for all $x>0$, $\log x < \sqrt{x}$. Thus, for any $X \sim P$, from (\ref{eqn:ent_cont}), we have
\[
|\hat{H}(X) - H(X)| \leq \sqrt{2|\calX| \delta(\hat{P},P)}.
\]
Using this and (\ref{eqn:emp_conc_dist}), the first inequality is obtained. Subsequently, using the triangle inequality and union bound, we obtain the convergence of the empirical mutual information.
\end{IEEEproof}

These rates of convergence are used to prove consistency.

\section{Proof of Lemma \ref{lemma:hypotheses_KL_bound}} \label{app:KL_bound_proof}
In this section we describe the proof of Lemma \ref{lemma:hypotheses_KL_bound}.

\begin{IEEEproof}
We first note that
\begin{align*}
D(Q_i \| Q_j) &= \Expect_{Q_i} \qth{\log \pth{\frac{Q_i(Y^\ell)}{Q_j(Y^\ell)}}} \\
&= \Expect_{Q_i} \qth{\log \pth{ \frac{ W_{Y_{i+1}Y_{i-1}} W_{Y_{j}Y_{j-1}} W_{Y_{j+1}Y_{j}} }{ W_{Y_{i+1}Y_{i}} W_{Y_{i}Y_{i-1}} W_{Y_{j+1}Y_{j-1}} } } } \\
&= \Expect_{Q_{\ell - i}}  \qth{\log \pth{ \frac{ W_{Y_{\ell-i+1}Y_{\ell-i-1}} W_{Y_{\ell-j}Y_{\ell-j-1}} W_{Y_{\ell-j+1}Y_{\ell-j}} }{ W_{Y_{\ell-i+1}Y_{\ell-i}} W_{Y_{\ell-i}Y_{\ell-i-1}} W_{Y_{\ell-j+1}Y_{\ell-j-1}} } } } \\
&= D(Q_{\ell-i}\|Q_{\ell-j}).
\end{align*}

Then we note that
\begin{align*}
D(Q_0 \| Q_1)  &= \sum_{y^\ell} \frac{1}{2} \prod_{i=2}^\ell W_{y_i y_{i-1}} \log \pth{2W_{y_2 y_1}} \\
&= 1 - h(1/2 - \epsilon),
\end{align*}
where $h(\cdot)$ is the binary entropy function. Similarly 
\[
D(Q_1 \| Q_0) = -\frac{1}{2} \log \pth{1-4\epsilon^2}.
\]

For any $1 < i < \ell$,
\begin{align*}
D(Q_0 \| Q_i) &= \Expect_{Q_0} \qth{\log \pth{\frac{2W_{Y_i Y_{i-1}}W_{Y_{i+1} Y_i}}{W_{Y_{i+1} Y_{i-1}}}}} \\
&= 1 + \pth{\frac{1}{2} + 2\epsilon - \epsilon^2} \log \pth{\frac{1}{2}+\epsilon} \\
&\quad + \pth{\frac{1}{2} - 2\epsilon + \epsilon^2} \log \pth{\frac{1}{2} - \epsilon}.
\end{align*}
Similarly, 
\[
D(Q_i \| Q_0) = -\pth{\frac{1}{2} - \epsilon} \log \pth{\frac{1}{2} + \epsilon} + \pth{\frac{1}{2} + \epsilon} \log \pth{\frac{1}{2} - \epsilon}.
\]

Having computed these distances, we make one additional observation. For $1 \leq i ,j \leq \ell$, $i \neq j$,
\begin{align*}
D(Q_i \| Q_j) &= \Expect_{Q_i} \qth{\log \pth{\frac{ W_{Y_{i+1}Y_{i-1}} } {W_{Y_{i+1}Y_{i}} W_{Y_{i}Y_{i-1}}} } + \log \pth{\frac{W_{Y_{j}Y_{j-1}} W_{Y_{j+1}Y_{j}}}{W_{Y_{j+1}Y_{j-1}}}} } \\
&= D(Q_0 \| Q_j) + D(Q_i \| Q_0).
\end{align*}

Since $\epsilon$ is bounded, $D(Q_0 \| Q_i) = O(1)$ and $D(Q_i \| Q_0) = O(1)$. This in turn proves that the KL divergences between any two hypotheses is a constant independent of $\ell$.
\end{IEEEproof}

\bibliographystyle{IEEEtran}
\bibliography{abrv,conf_abrv,lrv_lib}

\newcommand{\SortNoop}[1]{}
\begin{thebibliography}{10}
\providecommand{\url}[1]{#1}
\csname url@samestyle\endcsname
\providecommand{\newblock}{\relax}
\providecommand{\bibinfo}[2]{#2}
\providecommand{\BIBentrySTDinterwordspacing}{\spaceskip=0pt\relax}
\providecommand{\BIBentryALTinterwordstretchfactor}{4}
\providecommand{\BIBentryALTinterwordspacing}{\spaceskip=\fontdimen2\font plus
\BIBentryALTinterwordstretchfactor\fontdimen3\font minus
  \fontdimen4\font\relax}
\providecommand{\BIBforeignlanguage}[2]{{%
\expandafter\ifx\csname l@#1\endcsname\relax
\typeout{** WARNING: IEEEtran.bst: No hyphenation pattern has been}%
\typeout{** loaded for the language `#1'. Using the pattern for}%
\typeout{** the default language instead.}%
\else
\language=\csname l@#1\endcsname
\fi
#2}}
\providecommand{\BIBdecl}{\relax}
\BIBdecl

\bibitem{KitturCS2008}
A.~Kittur, E.~H. Chi, and B.~Suh, ``Crowdsourcing user studies with
  {M}echanical {T}urk,'' in \emph{Proc. SIGCHI Conf. Hum. Factors Comput. Syst.
  (CHI 2008)}, Apr. 2008, pp. 453--456.

\bibitem{IpeirotisPW2010}
P.~G. Ipeirotis, F.~Provost, and J.~Wang, ``Quality management on {A}mazon
  {M}echanical {T}urk,'' in \emph{Proc. ACM SIGKDD Workshop Human Comput.
  (HCOMP'10)}, Jul. 2010, pp. 64--67.

\bibitem{Scheiber2015}
N.~Scheiber, ``A middle ground between contract worker and employee,''
  \emph{The New York Times}, Dec. 2015.

\bibitem{GinoP2008}
F.~Gino and G.~Pisano, ``Toward a theory of behavioral operations,''
  \emph{Manuf. Service Oper. Manag.}, vol.~10, no.~4, pp. 676--691, Fall 2008.

\bibitem{KargerOS2014}
D.~R. Karger, S.~Oh, and D.~Shah, ``Budget-optimal task allocation for reliable
  crowdsourcing systems,'' \emph{Oper. Res.}, vol.~62, no.~1, pp. 1--24,
  Jan.-Feb. 2014.

\bibitem{MisraW2013}
V.~Misra and T.~Weissman, ``Unsupervised learning and universal
  communication,'' in \emph{Proc. 2013 IEEE Int. Symp. Inf. Theory}, Jul. 2013,
  pp. 261--265.

\bibitem{JungPL2014}
H.~J. Jung, Y.~Park, and M.~Lease, ``Predicting next label quality: A
  time-series model of crowdwork,'' in \emph{Proc. AAAI Conf. Human Comput. and
  Crowdsourcing (HCOMP'14)}, Nov. 2014, pp. 87--95.

\bibitem{KargerOS2013}
D.~R. Karger, S.~Oh, and D.~Shah, ``Efficient crowdsourcing for multi-class
  labeling,'' in \emph{Proc. ACM SIGMETRICS Int. Conf. Meas. Model. Comput.
  Syst.}, Jun. 2013, pp. 81--92.

\bibitem{ShahBW2016_arxiv}
N.~B. Shah, S.~Balakrishnan, and M.~J. Wainwright, ``A permutation-based model
  for crowd labeling: Optimal estimation and robustness,'' arXiv:1606.09632,
  Jun. 2016.

\bibitem{VempatyVV2014}
A.~Vempaty, L.~R. Varshney, and P.~K. Varshney, ``Reliable crowdsourcing for
  multi-class labeling using coding theory,'' \emph{{IEEE} J. Sel. Topics
  Signal Process.}, vol.~8, no.~4, pp. 667--679, Aug. 2014.

\bibitem{VarshneyJH2016}
L.~R. Varshney, P.~Jyothi, and M.~Hasegawa-Johnson, ``Language coverage for
  mismatched crowdsourcing,'' in \emph{Proc. 2016 Inf. Theory Appl. Workshop},
  Feb. 2016.

\bibitem{BanerjeeMDG2005}
A.~Banerjee, S.~Merugu, I.~S. Dhillon, and J.~Ghosh, ``Clustering with
  {B}regman divergences,'' \emph{J. Mach. Learn. Res.}, vol.~6, pp. 1705--1749,
  Oct. 2005.

\bibitem{Misra2015}
V.~Misra, ``Universal communication and clustering,'' Ph.D. dissertation,
  Stanford University, Jun. 2014.

\bibitem{ChanAEKL2015}
C.~Chan, A.~Al-Bashabsheh, J.~B. Ebrahimi, T.~Kaced, and T.~Liu, ``Multivariate
  mutual information inspired by secret-key agreement,'' \emph{Proc. {IEEE}},
  vol. 103, no.~10, pp. 1883--1913, Oct. 2015.

\bibitem{SlonimFT2002}
N.~Slonim, N.~Friedman, and N.~Tishby, ``Agglomerative multivariate information
  bottleneck,'' in \emph{Advances in Neural Information Processing Systems 14},
  T.~G. Dietterich, S.~Becker, and Z.~Ghahramani, Eds.\hskip 1em plus 0.5em
  minus 0.4em\relax Cambridge, MA: MIT Press, 2002, pp. 929--936.

\bibitem{ZhangSWW2016}
J.~Zhang, V.~S. Sheng, J.~Wu, and X.~Wu, ``Multi-class ground truth inference
  in crowdsourcing with clustering,'' \emph{{IEEE} Trans. Knowl. Data Eng.},
  vol.~28, no.~4, pp. 1080--1085, Apr. 2016.

\bibitem{NaganoKI2010}
K.~Nagano, Y.~Kawahara, and S.~Iwata, ``Minimum average cost clustering,'' in
  \emph{Advances in Neural Information Processing Systems 23}, J.~D. Lafferty,
  C.~K.~I. Williams, J.~Shawe-Taylor, R.~S. Zemel, and A.~Culotta, Eds.\hskip
  1em plus 0.5em minus 0.4em\relax MIT Press, 2010, pp. 1759--1767.

\bibitem{LiVVV_arxiv}
Q.~Li, A.~Vempaty, L.~R. Varshney, and P.~K. Varshney, ``Multi-object
  classification via crowdsourcing with a reject option,'' arXiv:1602.00575
  [cs.LG]., Jun. 2016.

\bibitem{AliS1966}
S.~M. Ali and S.~D. Silvey, ``A general class of coefficients of divergence of
  one distribution from another,'' \emph{J. R. Stat. Soc. Ser. B. Methodol.},
  vol.~28, no.~1, pp. 131--142, 1966.

\bibitem{Csizar1967}
I.~Csisz{\'{a}}r, ``Information-type measures of difference of probability
  distributions and indirect observations,'' \emph{Stud. Sci. Math. Hung.},
  vol.~2, pp. 299--318, 1967.

\bibitem{Dragomir2000}
S.~S. Dragomir, Ed., \emph{Inequalities for Csisz{\'{a}}r {$f$}-Divergence in
  Information Theory}, ser. RGMIA Monographs.\hskip 1em plus 0.5em minus
  0.4em\relax Victoria University, 2000.

\bibitem{Kailath1967}
T.~Kailath, ``The divergence and {B}hattacharyya distance measures in signal
  selection,'' \emph{{IEEE} Trans. Commun. Technol.}, vol. COM-15, no.~1, pp.
  52--60, Feb. 1967.

\bibitem{CsiszarZ2006}
I.~Csisz{\'{a}}r and Z.~Talata, ``Context tree estimation for not necessarily
  finite memory processes, via {BIC} and {MDL},'' \emph{{IEEE} Trans. Inf.
  Theory}, vol.~52, no.~3, pp. 1007--1016, Mar. 2006.

\bibitem{Yu1997}
B.~Yu, ``Assouad, {F}ano, and {L}e {C}am,'' in \emph{Festschrift for Lucien Le
  Cam: Research Papers in Probability and Statistics}, D.~Pollard,
  E.~Torgersen, and G.~L. Yang, Eds.\hskip 1em plus 0.5em minus 0.4em\relax New
  York: Springer, 1997, pp. 423--435.

\bibitem{ValiantV2011}
G.~Valiant and P.~Valiant, ``Estimating the unseen: An $n/\log(n)$-sample
  estimator for entropy and support size, shown optimal via new {CLTs},'' in
  \emph{Proc. 43rd Annu. ACM Symp. Theory Comput. (STOC'11)}, Jun. 2011, pp.
  685--694.

\bibitem{AntosK2001}
A.~Antos and I.~Kontoyiannis, ``Convergence properties of functional estimates
  for discrete distributions,'' \emph{Rand. Str. \& Alg.}, vol.~19, no. 3-4,
  pp. 163--193, 2001.

\bibitem{Paninski2003}
L.~Paninski, ``Estimation of entropy and mutual information,'' \emph{Neural
  Comput.}, vol.~15, no.~6, pp. 1191--1253, Jun. 2003.

\bibitem{NetrapalliBSS2010}
P.~Netrapalli, S.~Banerjee, S.~Sanghavi, and S.~Shakkottai, ``Greedy learning
  of {M}arkov network structure,'' in \emph{Proc. 48th Annu. Allerton Conf.
  Commun. Control Comput.}, Sep. 2010, pp. 1295--1302.

\end{thebibliography}

\end{document}